\documentclass[manuscript]{aastex}

\newcommand{\T}{\mathcal{T}}
\slugcomment{Submitted April 9, 2014}

\shorttitle{Magnetic Flux Ropes} \shortauthors{Hu et al.}

\begin{document}

\title{Structures of Interplanetary Magnetic Flux Ropes and Comparison with Their Solar Sources}

\author{Qiang Hu}
\affil{Department of Space Science/CSPAR, University of Alabama in
Huntsville, Huntsville, AL 35805}
%
%\author{C. D. Biemesderfer\altaffilmark{4,5}}
%\affil{National Optical Astronomy Observatories, Tucson, AZ 85719}
\email{qh0001@uah.edu}

\author{Jiong Qiu}%\altaffilmark{5}}
\affil{Department of Physics, Montana State University, Bozeman,
MT 59717-3840} \email{qiu@physics.montana.edu}

\and

\author{B. Dasgupta, A. Khare\altaffilmark{1}, and G.~M. Webb}
\affil{Center for Space Plasma and Aeronomic Research (CSPAR),
University of Alabama in Huntsville, Huntsville, AL 35805}
%\email{bd0001@uah.edu}

\altaffiltext{1}{Department of Physics and Astrophysics,
University of Delhi, Delhi, 110007, India}

%\author{A. Khare}%\altaffilmark{5}}
%\affil{Center for Space Plasma and Aeronomic Research (CSPAR),
%University of Alabama in Huntsville, Huntsville, AL 35805}
%\email{ak0005@uah.edu}
%
%\and
%
%\author{G. M. Webb}%\altaffilmark{5}}
%\affil{Center for Space Plasma and Aeronomic Research (CSPAR),
%University of Alabama in Huntsville, Huntsville, AL 35805}
%\email{gmw0002@uah.edu}

\begin{abstract}
%Whether a magnetic flux rope is pre-existing or in-situ formed in
%the Sun's atmosphere, there is little doubt that
Magnetic reconnection is essential to release the flux rope during
its ejection. The question remains: how does
the magnetic reconnection change the flux rope structure?
Following the original study of \citet{Qiu2007}, we compare properties of ICME/MC
flux ropes measured at 1 AU and properties of associated solar
progenitors including flares, filaments, and CMEs. In particular,
the magnetic field-line twist distribution within interplanetary
magnetic flux ropes is systematically derived and examined. Our
analysis shows that for most of these events, the amount of
twisted flux per AU in MCs is comparable with the total
reconnection flux on the Sun, and the sign of the MC helicity is
consistent with the sign of helicity of the solar source region
judged from the geometry of post-flare loops. Remarkably, we find
that about one half of the 18 magnetic flux ropes, most of them
being associated with erupting filaments, have a nearly uniform
and relatively low twist distribution from the axis to the edge,
and the majority of the other flux ropes exhibit very high twist
near the axis, of up to $\gtrsim 5$ turns per AU, which decreases
toward the edge. The flux ropes are therefore not linear force
free. We also conduct detailed case studies showing the contrast
of two events with distinct twist distribution in MCs as well as
different flare and dimming characteristics in solar source
regions, and discuss how reconnection geometry reflected in flare
morphology may be related to the structure of the flux rope formed
on the Sun.
\end{abstract}

\section{INTRODUCTION}
Observations of Magnetic Clouds (MCs) obtained in-situ by various
spacecraft missions provide the most direct and definitive
evidence for the existence of magnetic flux ropes that originate
from the Sun. Despite the debate on formation mechanisms of such
flux ropes, it is acknowledged that, for a
flux rope to erupt out of the Sun, magnetic reconnection has to be
invoked. Magnetic reconnection allows a change of connectivity
between different magnetic domains, or the magnetic topology.
Through this change, the magnetic shear created by turbulent
plasma motion in or below the photosphere is transferred to a
twisted magnetic structure, such as a flux rope, which is then
ejected from the Sun \citep{Low1996,Demoulin2006} often in the
form of a Coronal Mass Ejection (CME). On many grounds, reconnection
on the Sun is a viable mechanism for the formation of
flux rope structure as well as its energetics during its
evolution near the Sun; however, it has been a tremendous
challenge to observationally establish an unambiguous and
quantitative association between flux rope properties and relevant
magnetic reconnection properties.

We have been able to measure previously the magnetic  reconnection
flux during flares in comparison with the flux budget of magnetic
clouds observed a few days after the flare/CME eruption
\citep{Qiu2007}. The study, though with a relatively small sample
of 9 events, showed that the total reconnection flux during a
flare, spanning two orders of magnitudes in these events, is
comparable with the amount of twisted magnetic flux in the
associated MCs, suggesting that these flux ropes are likely to
have been formed by reconnection in the corona in the wake of its
eruption. Apart from the total reconnection flux, morphology
evolution of flares may also provide information on reconnection
geometry and the resultant flux rope structure. To form the flux
rope, theoretical models have envisaged a certain sequence of
magnetic reconnection. For example, observations have shown that
reconnection in the early stage forms post-flare loops highly
sheared relative to the magnetic polarity inversion line (PIL),
and then ribbons expand in a direction perpendicular to the PIL
forming less sheared post-flare loops
\citep{Moore2001,Fletcher2004,Su2007,Qiu2010,Cheng2012}. These
observations are qualitatively consistent with models of flux rope
formation as depicted by \citet{Ballegooijen1989}, and more
recently by \citet{Aulanier2010,Aulanier2012}, which would predict
that the flux rope is less twisted near its axis and more twisted
further out. Alternatively, \citet{Longcope2007a} illustrates a
scenario of sequential reconnection between a flux rope in the
making and a sheared arcade, which starts from one end of the rope
axis and progresses to the other end. Such continuous reconnection
produces a highly twisted flux rope. The model predicts that flare
ribbons are not brightened simultaneously but instead sequentially
along the PIL, which may be evident in observations of many
two-ribbon flares exhibiting the so-called ``zipper effect", such
as the famous Bastille-day flare \citep[][and references
therein]{Qiu2010}. Being able to infer reconnection properties by
observing flare signatures on the Sun's surface therefore provides
information to help distinguish these different models, and
predict the structure of the infant flux rope that is formed by
reconnection \citep{Longcope2007b,Qiu2009}.

Having formed on the Sun, the magnetic structure of flux ropes has
been exclusively derived from in-situ measurements a few days
after their ejection toward the observer. There has been a
continuous effort in modeling flux-rope structures embedded within
the Interplanetary CME (ICME) complex, utilizing in-situ
spacecraft measurements across such structures. These models range
from a traditional one-dimensional (1D) configuration to a fully
two-dimensional (2D) model of the Grad-Shafranov (GS)
reconstruction. We employ the GS method here to examine the
flux-rope structures for more than a dozen ICME events by
utilizing in-situ measurements from spacecraft ACE, Wind and
STEREO. In particular, we systematically derive the magnetic
field-line twist distributions within the core regions for the
events that exhibit a typical flux-rope configuration based on GS
reconstruction results. The study of field-line twist within flux
ropes had been reported before for individual events
\citep{2009SoPhM,2009JGRAM,2008ApJLL,2002JGRAHu} and with
different approaches \citep[e.g.,][]{2006AandAD}, but not in the
systematic and congregated manner that we will report here. The
twist of magnetic flux ropes is closely related to the field-line
lengths within the ropes. Theoretically they are all dependent on
models utilized in the analysis of in-situ data.
\citet{Larson1997} presented the first study of energetic electron
transit timing observations between the electron release on the
Sun and arrival at 1 AU  to derive the field-line length directly
for one event. That study provided support for the
linear-force-free flux-rope model of MCs.
\citet{2011JGRAK,2011ApJK} recently extended that study by
examining more events, utilizing the same date sets from the Wind
spacecraft and additional measurements from ACE, following a
similar approach. They showed the comparison of field-line length
measurements with certain theoretical flux-rope models and the
general inapplicability of a linear force-free field model. Such a
model possesses a field-line length (and twist) distribution that
increases with radius at a greater rate than that derived from
electron onset observations \citep{2011JGRAK}. However comparison
with the corresponding GS reconstruction results showed improved
consistency and will be reported in a separate paper. In this
paper, we will employ the GS reconstruction method to analyze MC
observations and measure the twist distribution in MCs. We present
a detailed description of the methodology and a quantitative
analysis of magnetic field-line twist. Moreover we carry out
additional studies to connect with their solar source regions and
offer interpretations of such connections.

In this investigation, we strive to examine the role of magnetic
reconnection in the formation and evolution of magnetic flux ropes
in the corona by relating the in-situ analysis results to the
corresponding solar source regions in a quantitative manner. We
recognize that such an approach cannot provide direct and
deterministic evidence for the formation process of flux ropes,
because flux ropes are {\em magnetically invisible} on the Sun and
further out in low corona. The present observations of commonly
recognized plasma structures in flux ropes on the Sun, including
filaments, sigmoids, erupting loops/arcades, and CMEs observed in
a variety of wavelengths, are still a large step away from being
able to yield a close estimate of the amount and distribution of
twist in these structures \citep[see review by][]{Vourlidas2014}.
Measurements of reconnection flux from flare observations,
alternatively, allow us to indirectly {\em infer} magnetic
properties that can be related to flux rope formation and
evolution. Direct measurements of magnetic properties of flux
ropes have been nearly exclusively derived from in-situ
observations, and there is a large gap, namely the interplanetary
space of distance 1AU starting from the Sun's corona, between
these two kinds of observations. Nevertheless, it is hoped that
large-scale numerical models can make a crucial link with valid
observational constraints at the two ends that we attempt to
provide here, and in this process, elucidate the physical
mechanisms governing formation and evolution of magnetic flux
ropes \citep[e.g.,][]{Fan2010,Karpen2012,Aulanier2012,Titov2012}.

In this paper, we use an enlarged sample of 19 events observed
from 1998 through 2011 by a variety of instruments, the latest
being SDO and STEREO, with identified association between MCs and
solar progenitors including CMEs, flares, and filament eruptions.
The comprehensive information of these events is given in
Table~\ref{info_list}. {{Identification of these events will be
discussed in the next section.}} Note that whereas our previous
research focused on events with major flares and fast CMEs, this
enlarged sample includes events associated with
filament/prominence eruptions (P.E.) from the quiet Sun without
major flares. From the table, it is also seen that a number of
these events are associated with slow or moderate CMEs. For some
of the more recent events, observations by both SDO/AIA and STEREO
are available and examined, allowing us to conduct more detailed
case studies of flares observed on the disk by AIA and CMEs
observed by STEREO. We discuss identification of these events in
Section 2, and present methods of flux rope modeling in Section 3,
analysis of solar observations in Section 4, summary and
comparison of these measurements in Section 5, followed by
conclusions and discussions in the last section.

\section{IDENTIFICATION OF MC, CME, AND SOLAR SURFACE ACTIVITIES}
{For meaningful comparison between properties of flux ropes
observed at 1~AU and their solar progenitors, identification of
MC, CME, and associated solar surface activities is crucial. Among
the 19 events studied in this paper, the first 9 events are
samples in our previous work \citep{Qiu2007}. These events
occurred between 1998 and 2005, and the association between MCs,
CMEs, and solar surface activities was identified by seven
different groups listed as references in Table 1 of
\citet{Qiu2007}, aided with authors' own examination of flare and
CME observations by a cluster of instruments including LASCO, EIT,
TRACE, and Big Bear Solar Observatory. The other events, except
events \#16 and 17 in Table~1, are selected from
\citet[][hereafter referred as LI catalog]{Li2014}. Event \# 16 is
selected from \citet{Gopal2012}, and event \#17 is identified
through private discussion with Dr. C. C. Wu. These events (\#10 -
19) occurred during 2008 through 2011, when CMEs and ICMEs can be
observed and tracked in STEREO observations while associated solar
surface activities are observed by AIA onboard SDO (except event
\#10). For identification, \citet{Li2014} searched ``the LASCO CME
catalog for halo or partial halo CMEs during the five days prior
to the MC arrival" and also used ``STEREO coronagraph and HI
images for better certainty of the correspondence." Most of these
events (\#10 - 19) are also found in two other catalogs compiled
by Phillip Hess and Jie Zhang at
$http://solar.gmu.edu/heliophysics/index.php/GMU\_CME/ICME\_List$
(abbreviated as HZ hereafter) and by I. Richardson and H. Cane at
$http://www.srl.caltech.edu/ACE/ASC/DATA/level3/icmetable2.htm$
(denoted as RC hereafter). The LI and HZ catalogs identify CMEs as
well as times and locations of flares or filaments associated with
ICMEs, and the RC catalog only lists CMEs associated with ICMEs.
Some of these events have also been analyzed, modeled, and
reported in published literature. These references are provided in
Table~\ref{info_list}. In the LI, HZ, and RC catalogs, magnetic
clouds are identified from ACE observations, and CME information
is given according to LASCO observations. In some other references
such as \citet{Mostl2014} and \citet{Harrison2012}, CMEs are also
tracked in STEREO observations all the way to 1~AU, and arrival
times at STEREO and Wind spacecraft are estimated and compared
with observations.}

We do notice that these references do not agree on the identification of solar sources for
a few events, and for these cases, we adopt the association recognized by the majority of
these authors. Below we discuss details of event identification by different sources that can
be found in public literature.

For event \#11, LI, \citet{Lugaz2012}, and \citet{Mostl2014} all
identified the MC on 2010 May 28 to be associated with the CME at
18:30 (LASCO C2) on 2010 May 23, and LI and \citet{Lugaz2012} both
recognized the association with an erupting filament and B1 flare
on the Sun's disk. Note that the disk location of the
flare/filament event is N19W12, as reported by \citet{Lugaz2012}
and confirmed by our own scrutiny (see Figure~\ref{fig0528_1}),
different from the location N20E10 reported in LI. In HZ and RC
catalogs, however, the MC is considered to be associated with a
CME at 14:06 UT (LASCO C2) on 2010 May 24. \citet{Lugaz2012} have
analyzed and modeled this event, showing that the CME on May 24
caught up with the one on May 23, and the CME that occurred a day
later was deflected whereas the CME on May 23 reached L1 to be
observed by Wind. We therefore consider the association between
the MC on May 28 and CME/flare/filament on May 23 to be reliable.

For event \#12, numerous research groups have reported analysis
and modeling of the CME/flare/filament events on 2010 August 1
possibly associated with the MC on August 4. Association with a
C3.2 flare at 07:32 UT is reported in both LI and HZ catalogs;
however, the three catalogs HZ, RC, and LI, identify three
different LASCO CMEs, taking place at 03:54~UT, 09~UT (also
identified by \citet{Mostl2014}), and 13:42~UT, respectively, to
be associated with the MC. On 2010 August 1, three filament
eruptions were observed roughly at 3~UT, 8~UT, and 18~UT by AIA
and STEREO \citep[e.g.,][ and other references listed in
Table~\ref{info_list}]{Schrijver2011,Torok2011,Titov2012}. By
studying the STEREO images, \citet{Harrison2012} further
identified four CMEs with reconstructed onset times at 3~UT, 8~UT,
10~UT, and 16~UT, three of them (at 3, 10, and 16~UT) being
associated with three different filament eruptions, and the one at
8~UT being associated with the C3.2 flare \citep[also
see][]{Temmer2012}. \citet{Harrison2012} also predicted the
arrival times of 3 CMEs (at 8, 10, and 16 UT) at Wind spacecraft
to be August 3 12~UT, August 4 8~UT and 16~UT, respectively. If
identification by \citet{Harrison2012} is accurate, the MC
analyzed in this paper is likely related with either the 8~UT CME
with a flare, or the 10~UT CME with a filament eruption. Note that
the flare and filament eruption, though close in time, occurred in
two different active regions. Furthermore, CMEs launched at
different times throughout the day probably interacted with each
other \citep[e.g.,][]{Harrison2012,Mostl2012,Temmer2012}.
Therefore, there is a great difficulty to find an unambiguous
one-to-one association between the MC and flare/CME/filament. In
this paper, we still report properties measured in the C3.2 flare,
which is the only major flare on this day and is most likely
associated with the CME at 8~UT
\citep[STEREO;][]{Harrison2012,Mostl2014} or 9~UT (LASCO; RC), but
with the caution that a direct comparison between flare and MC
properties is not entirely justified for this event before fully
understanding the relationship between all the different events
occurring on the same day.

Event \#14 is found in LI and HZ catalogs. The MC on 2011 March 29
is identified to be associated with a LASCO CME at 14:36~UT on
March 25 in LI, but is thought to be related with a LASCO CME at
02:00~UT on March 25 in HZ.
%(the CME timing is given as 05:36~UT in HZ, which should be a typo since it cannot be found in the LASCO CME catalog).
Tracking the event in STEREO EUVI, COR1, COR2, and HI images, \citet{Savani2013} identified the MC to be associated
with a CME that entered the STEREO COR-2A view at 21:24 UT on March 24. In terms of solar surface activities,
both LI and HZ catalogs register a C1.0 flare in an active region located at S16E31. It appears that four flares
(C1.4 at 20:53 on March 24, C1.0 at 00:57, C1.0 at 16:47, and M1.0 at 23:08 on March 25) took place in this
same active region around the times of the above-identified CMEs. Some of these flares or CMEs are
also associated with filament eruptions. For the very large ambiguity in
identifying associated CME and flare or filament, as reflected in the disagreement among the above references,
we cannot determine solar source properties for this event. However, since all flares or filament eruptions,
which are probably candidates of the MC source, occur in the same active region, the morphology
of the flares in the active region allows us to determine the sign of the helicity (see Section 4). Furthermore,
this active region produces small flares, the biggest one being the M1.0 flare. The reconnection flux measured
in this largest flare and reported in Table~\ref{table2} serves as an upper limit of reconnected flux, if any,
associated with the MC flux rope.

For event \#16, the association between MC, CME, and flare is
identified by \citet{Gopal2012}, and the same association is also
confirmed in HZ and RC catalogs. The identification is therefore
regarded to be unambiguous.

The MC of event \#17 was best observed as well as measured in
STEREO A. It is identified to be associated with a CME and a limb
flare, without filament eruption, through private discussion with
C. C. Wu who modeled  this event, as well as by authors' own
examination of AIA, STEREO, and LASCO movies.

Event \#18 is reported in all three catalogs LI, HZ, and RC, in
all of which the MC is associated with the CME at 0:05~UT on
September 14. LI identifies a C2.9 flare in active region 11289 at
N23W21 (see Figure~\ref{fig0914_1}) to be associated with the
CME/MC; HZ also identifies the solar source to be in the same
active region at the same location, though without listing a flare
in the catalog. For the general agreement among the above three
catalogs, identification of this event is also regarded to be
reliable.

Event \#19 is reported in two catalogs LI and HZ, as well as by
\citet{Mostl2014}. In LI, the MC is identified to be associated
with a LASCO CME at 01:25~UT on 2011 October 22 and a filament
eruption at N30W30, which did not produce an obvious flare.
\citet{Mostl2014} associated the MC with a CME seen in STEREO
COR-2 at 1:09~UT. However, HZ identifies the MC to be associated
with a LASCO CME at 10:36~UT and an M1 limb flare peaking at
11:10~UT in a different region at N25W77. For close proximity
between LI and \citet{Mostl2014}, aided with authors' own
examination of the AIA and STEREO movies, we adopt the
identification by LI for this event.

In summary, to our best knowledge and based on available published
literature including on-line catalogs, identification of MCs and
their solar sources is reliable in most of the events listed in
Table~\ref{info_list}. There is a large uncertainty in event \#14,
limiting our MC/flare comparison to being only qualitative. The
complex nature of event \#12 does not allow us to establish a
one-to-one relation between the MC and its solar source. We still
report measurements for these two events for reference. For the
rest of the events, we measure and compare properties of MCs and
their solar sources.

\section{GRAD-SHAFRANOV RECONSTRUCTION OF MAGNETIC FLUX ROPES}
The structures of magnetic flux ropes embedded within ICMEs
propagated from the low corona and detected in-situ by spacecraft ACE,
Wind and STEREO etc. are examined by the Grad-Shafranov (GS)
reconstruction method
\citep{2001GeoRLH,2002JGRAHu,2006JGRAS,2013SoPhH}. The GS method
is a truly two-dimensional (2D) method that yields a solution to
the Cartesian GS equation describing a 2$\frac{1}{2}$D magnetic
field, utilizing the spacecraft measurements of
both the magnetic field and bulk plasma parameters across the
structure along a single path.

\subsection{General Approach and Output} \label{GSmthd}
The general approach of GS reconstruction is based on a
cylindrical geometry with the $z$-axis being the flux-rope axis of
translation symmetry such that $\partial/\partial z\approx 0$. The
transverse plane ($x,y$) is perpendicular to $z$ and
the GS equation governing the plasma structure in
quasi-static equilibrium is
\citep{Sturrock1994,1999JGRH}
\begin{equation}\frac{\partial^2 A}{\partial x^2}+\frac{\partial^2 A}{\partial
y^2}=-\mu_0\frac{dP_t}{dA}=-\mu_0
j_z(A),\label{eq:GS}\end{equation} where a magnetic flux function
$A$ is defined such that the transverse magnetic field components
are $B_x=\partial A/\partial y$, and $B_y=-\partial A/\partial x$.
The equi-value contours of $A$
represent transverse magnetic field lines. Therefore the
transverse field on the cross-section of a flux rope is completely
determined by the scalar flux function $A(x,y)$ and the magnetic
poloidal flux is directly calculated by taking the difference of
the $A$ values between  two iso-surfaces of $A$, then multiplied
by a chosen length $L$ along the $z$ axis \citep{Qiu2007}. These
iso-surfaces of $A$ are nested distinct cylindrical surfaces,
called $A$ shells, on which the magnetic field lines are winding
along the $z$ axis.

The other important quantity is the so-called transverse pressure
$P_t(A)$ that appears on the right-hand side of the GS
equations~(\ref{eq:GS}) and is a single-variable function of $A$.
Its first-order derivative yields the axial current density
$j_z(A)$. This function is the sum of the plasma pressure $p$ and
the axial magnetic pressure $B_z^2(A)/2\mu_0$. Both are functions
of $A$ alone. This important feature allows us to devise an
algorithm for determining the invariant $z$ axis, in turn checking
for the validity of the translation symmetry and finally obtaining
the axial field distribution over the solution domain once the GS
equation~(\ref{eq:GS}) is solved to obtain a solution $A(x,y)$
within a rectangular domain. Detailed description of the
procedures was given in prior works \citep[see,
e.g.,][]{2002JGRAHu}. A quantitative measure $R_f$ that evaluates
the goodness-of-fit of spacecraft data to a functional form
$P_t(A)$ was defined in the last few steps of the GS
reconstruction to assess, partially, the quality of the
reconstruction results \citep{Hu2004}.

\begin{figure}
\begin{minipage}[b]{.6\textwidth}
\includegraphics[width=1.\textwidth]{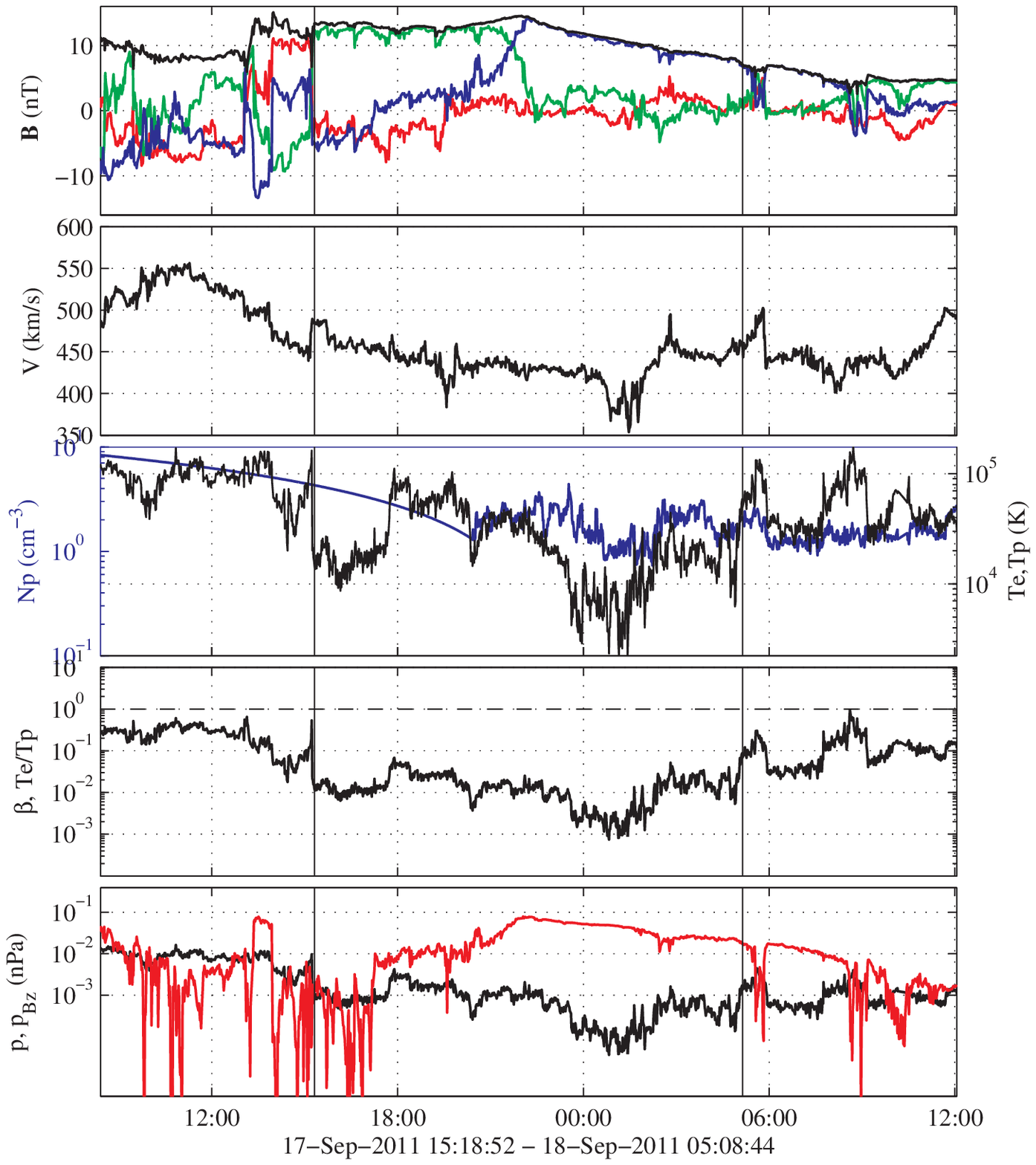}
\centering (a)
\end{minipage}
%\caption{} \label{fig0914_data}
%\end{figure}
%\begin{figure}
\begin{minipage}[b]{.4\textwidth}\centering (b)
\includegraphics[width=.8\textwidth,clip=]{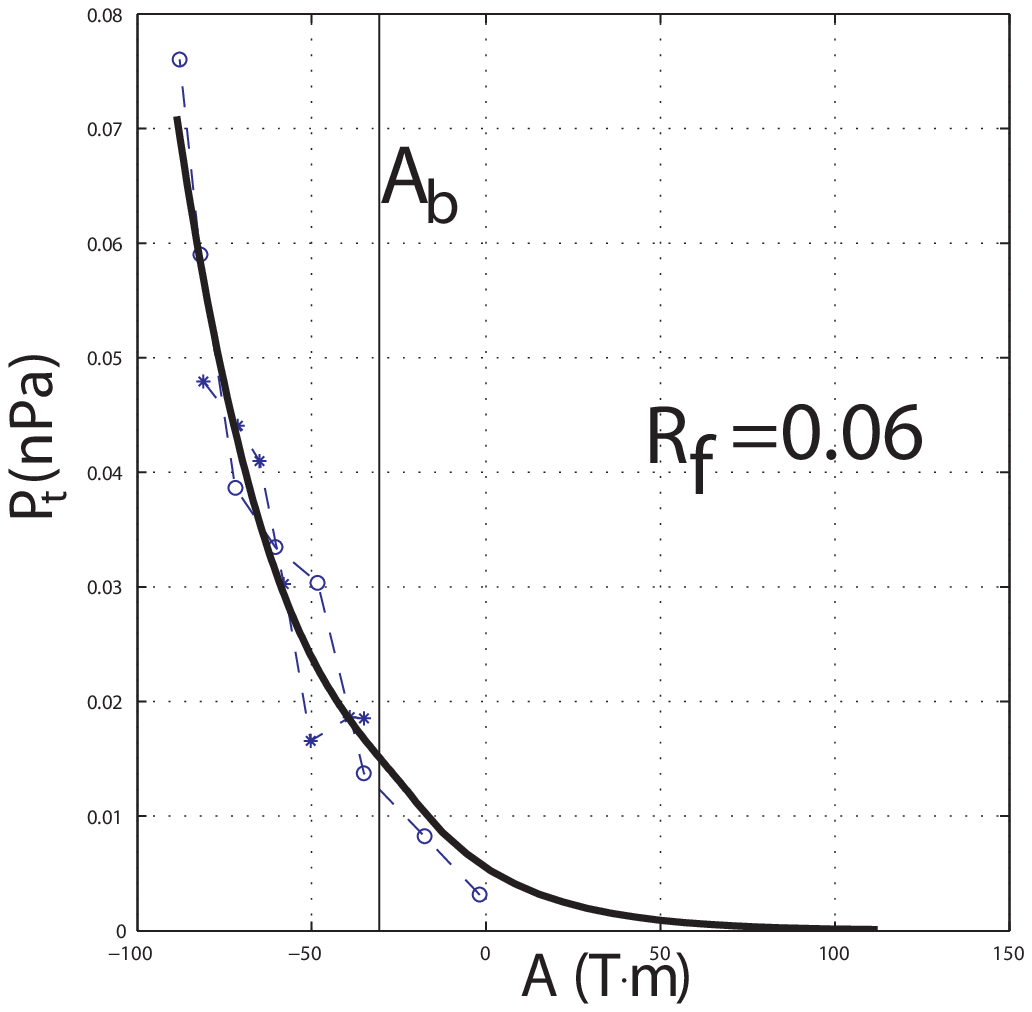}
\includegraphics[width=1.\textwidth]{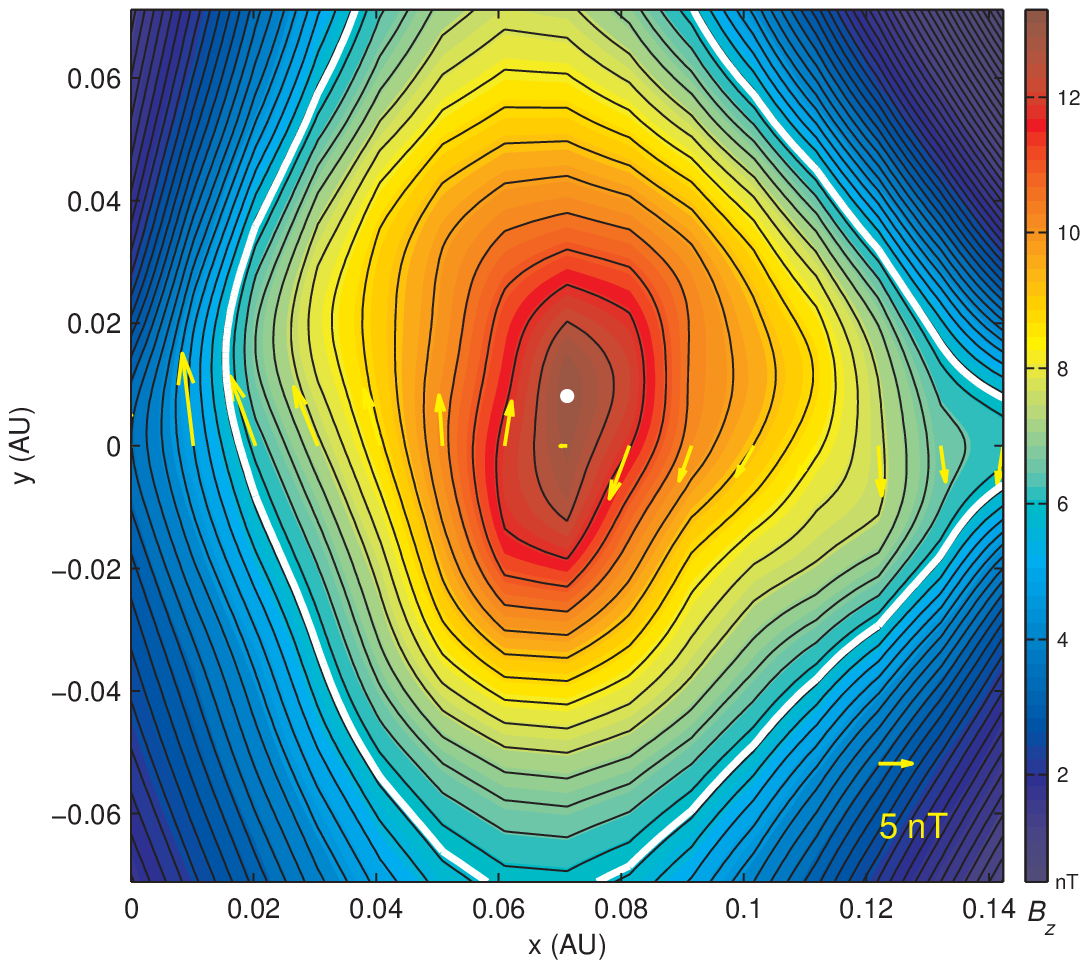}
(c)
\end{minipage}
%\vspace{-1.\textwidth}   % Shift close to the panel top
%     \centerline{%\Large \bf     % Includes the labels (here needs the color package)
%      %\hspace{0.04 \textwidth} \color{white}{(a)}
%      \hspace{0.92\textwidth}  {(b)}
%         \hfill}
%     \vspace{1.\textwidth}
\caption{An example of basic GS reconstruction result for event
\#18 in Table~\ref{info_list}. (a) Time series of ACE spacecraft
measurements: (from top to bottom panels) the in-situ magnetic
field magnitude (black) and GSE-X (red), Y (green), and Z (blue)
components, the plasma bulk flow speed, the proton density (left
axis; blue) and proton and electron (if available) temperature
(right axis), the plasma $\beta$ and electron over proton
temperature ratio (if available), and the plasma and axial
magnetic field (red) pressure.  The vertical lines mark the GS
reconstruction interval as given beneath the last panel. (b) The
measurements of $P_t(x,0)$ versus $A(x,0)$ and the fitted $P_t(A)$
curve (thick black line). The flux rope boundary is marked at
$A=A_b$ and a fitting residue $R_f$ is denoted. (c) The
cross-sectional map of the solution $A(x,y)$ (black contour lines)
and the axial field $B_z(A)$ (filled  contours in color). The
yellow arrows are the measured transverse magnetic field along the
spacecraft path ($y=0$). The white contour line denotes the
boundary $A=A_b$ while the white dot denotes the center where the
axial field is the maximum and $A\equiv A_0$.} \label{fig0914_ALL}
\end{figure}

An example of the basic GS reconstruction results is given in
Figure~\ref{fig0914_ALL} for event \#18 in Table~\ref{info_list}.
Figure~\ref{fig0914_ALL}a shows the time series of in-situ ACE
spacecraft measurements of the ICME event on 17 September 2011,
from which both the magnetic field vectors and plasma density,
temperature (including electron temperature $T_e$ if available)
and velocity were utilized in generating the GS reconstruction
results. The interval marked by two vertical lines is the GS
interval given in Table~\ref{info_list} and was chosen for the
analysis. It clearly corresponds to a region of low proton $\beta$
value in this case. In particular, the total plasma pressure and
the axial magnetic pressure as plotted in the bottom panel
indicate a region dominated by the magnetic field during the GS
interval. For this event, no $T_e$ data were included. Based on
the recent study of \citet{2013SoPhH}, the inclusion of $T_e$
generally has a negligible effect on the topological properties of
the results, such as axis orientation, or the size and shape of
the cross-section, but there is a 10-20\% discrepancy in other
physical quantities. Figure~\ref{fig0914_ALL}b shows the plot of
$P_t$ versus $A$ typical of a flux-rope solution. The scattered
symbols are measurements while the thick curve represents an
analytic functional fit of $P_t(A)$ to the data points. A fitting
residue, $R_f$, is calculated to show the quality of the fit, the
smaller the $R_f$ value, the more reliable the overall
reconstruction results. The rule of thumb is that in general a
value not exceeding 0.20 is considered acceptable. A boundary
value $A=A_b$ is defined and marked such that the GS solution of
the flux-rope configuration is most valid within this boundary
($A< A_b$ in this case), as also highlighted by the thick white
line in panel (c). Figure~\ref{fig0914_ALL}c shows a typical
presentation of the GS solution on the cross-sectional ($x-y$)
plane which represents a cut of the cylindrical structure
perpendicular to the $z$ axis. The concentric contour lines
represent the transverse field lines while the color-filled
contours are the axial field distribution with scales indicated by
the colorbar to the right. Therefore this shows the full
characterization of the three-component magnetic field within the
solution domain. This cross-sectional map shows a flux-rope
solution with left-handed chirality with closed loops surrounding
the center that was crossed by the spacecraft in close vicinity
(the spacecraft path is always along $y=0$).

\subsection{Magnetic Field-line Twist}\label{sectau}
To further visualize the GS reconstruction result and facilitate
detailed analysis of magnetic field-line twist and length (the
latter to be reported elsewhere), we present a 3D view of the
flux-rope solution by drawing selected field lines in a 3D volume
extended along the $z$ axis in Figure~\ref{fig0914_3D}. So the
cross-sectional map of Figure~\ref{fig0914_ALL}c corresponds to a
projection of these spiral field lines as viewed along the $z$
axis. Therefore only the field lines completing at least one full
turn will appear as closed loops in Figure~\ref{fig0914_ALL}c. We
denote the outermost loop with corresponding value
$A'=|A-A_0|\equiv A_c$. In Figure~\ref{fig0914_3D}, only three
representative field lines are drawn, one near the center (red)
and the other two on outer loops, but are all within $A'=A_c$.
Therefore the one near the center appears straight and the other
two appear to be winding along the $z$ axis with distinct twist.
The field-line twist can be evaluated from these graphic
representation based on the reconstructed magnetic field vectors
in the volume. For example, for each of the blue and pink outer
field lines, the root on the $z=0$ plane is denoted by a circle,
and the field line can be traced from the root point in the
volume.  The point along the field line at which one full turn is
completed is marked by a cross. If we denote the length along $z$
dimension between the circle and the cross by $L_z$ in AU, then
the twist for that particular field line is
\begin{equation}\tau(A)=\frac{1}{L_z},\end{equation} in unit of
turns/AU. This procedure can be done for all points rooted on that
particular loop at $z=0$ plane of the same $A$ value, i.e., by
moving the circle around the same loop. Apparently all these field
lines should have the same $L_z$ value, thus the twist $\tau$ is a
function of $A$ alone. We repeat these
procedures for all root points on all closed loops to obtain an
estimate of $\tau$ and associated uncertainty as  a function of
$A$. A few other methods of approximating the field-line twist for
cylindrical flux ropes are described in the Appendix. Detailed
studies and validation of these methods are presented there for a
few analytic flux-rope models whose field-line twist distributions
are exactly known. The test case studies show that the graphic
method described here yields the most reliable estimate of
magnetic field-line twist, and will be utilized primarily in
analyzing the events to be presented. Our interpretations will
also be based on the results obtained from this method.

\subsection{Summary of GS Reconstruction Results}
Various physical quantities have been derived from
the GS reconstruction output. These include the axial field
$B_z$, the toroidal (axial) and poloidal magnetic flux, $\Phi_t$
and $\Phi_p$, the relative magnetic helicity $K_r$, the axial
current density $j_z$ and current $I_z$, and the field-line twist.
They all can be calculated and presented as functions of $A$, as
discussed in Sections~\ref{GSmthd} and \ref{sectau}. Together
with the field-line twist estimates, we systematically present the
distribution of these quantities along $A$ shells for events
\#1-19, except for event \#13 for which the GS reconstruction
results are not available.

Figure~\ref{figALL} shows a summary plot of the distributions of
all the aforementioned quantities versus the shifted flux function
$A'=|A-A_0|$. The integral quantities such as magnetic flux,
helicity and current increase monotonically with $A'$ since their
distributions represent accumulative sums over increasing area or
volume across the $A$ shells from the center ($A'\equiv0$) to the
boundary of the flux rope. The axial magnetic field also shows
monotonically deceasing behavior typical for such flux-rope
structures. The axial current density, on the other hand, shows
the greatest variation since it represents the first-order
derivative of $P_t(A)$ along the $A$ shells. The field-line twist,
as given here from two approximate methods only for illustrative
purpose, ranges from about 2 to 20 turns/AU. They show a general
trend of rapid decreasing from the center or constant twist and
the smaller the size of the flux rope is, the larger the twist
number becomes. Additional and more reliable results from the
graphic method will be presented and discussed below. We further
separate the summary plot of Figure~\ref{figALL} into two subplots
in Figure~\ref{figVAR} corresponding to the events associated with
P.E. and without P.E., respectively.  The two groups show a slight
distinction between them. On average, MCs not associated with P.E.
appear to carry slightly larger twist than those associated with
P.E.. We also note a prominent non-P.E. event (\#16) of small size
and the greatest twist number. Such a profile, although extreme
(note the GS interval duration is the shortest, about 2 hrs), is
reliable since all three estimates of the field-line twist
(especially $\tau_{dF}$) agree with the twist obtained from the
graphic method.

Table~\ref{table2} summarizes some of the  results, especially the
total magnetic flux and helicity content within the flux-rope
boundary $A=A_b$ (denoted with the additional subscript ``$max$").
The  corresponding estimates of the average twist within such a
boundary are calculated as $\bar\tau_H$ and $\bar\tau_F$ according
to equations~\ref{eq:tauH} and \ref{eq:tauF}, for references
purposes. They seem to compare well with the averages and standard
deviations of $\tau(A)$ listed in the last column. The axial
magnetic field at the flux rope center $B_{z0}$ and the helicity
sign are also given, together with the helicity sign of the solar
source region and the reconnection flux $\Phi_r$ to be described
in Section~\ref{sec:flux}. The helicity signs agree well with a
13/14 match rate, excluding events marked with ``U". Detailed
comparisons among these quantities will be discussed in
Section~\ref{sec:flux}.

The last column of Table~\ref{table2} gives the average and
standard deviation of the twist distribution along the $A$ shells
obtained by the graphic method, as displayed in
Figure~\ref{figtauA} with uncertainties. Figure~\ref{figtauA}a
shows the variation of $\tau$ along  the closed $A$ shells (loops)
as a function of the shifted flux function $A'$ for all events
such that the center of the flux rope always corresponds to
$A'\equiv 0$. All the lines extend from the center to the
outermost loop of $A'=A_c$ which differs for different events and
can be regarded as a proxy for the transverse size of the
flux-rope structure. The associated error bars are small
(generally less than the thickness of each line), indicating
excellent determination of twist by the graphic method. The
overall trend is that the twist either largely decreases rapidly
from the center or remains small and fairly constant throughout
the flux-rope structure. The twist values range from a little
above 1.5 to about 25 turns/AU, and the smaller the size of the
flux rope is, the larger the twist becomes. There is no clear
indication of significant increase of twist with increasing $A'$.
We further separate our events into two groups based on their
association with or without prominence eruption, and present the
results in panels (b) and (c), respectively. They show the same
general trend as panel (a) and appear to have no drastic
distinctions in size and twist value characteristics. The event of
the greatest twist value and a monotonically decreasing gradient
with respect to $A'$ is a non-P.E. event (\#16).

Figure~\ref{figtauAc} visualizes the results in the last column of
Table~\ref{table2}. Here the vertical bars represent the standard
deviations of $\tau(A)$ for each event (corresponding to each line
in Figure~\ref{figtauA}a), indicating the degree of variation of
$\tau$ within each flux-rope structure. The plot reinforces the
pattern of smaller the size, larger the twist and twist variation.
Most P.E. associated events show little variation with small
vertical bars, while some non-P.E. events show significant
variations. As we will demonstrate in the case studies, such
variations are indications of a strong gradient in field-line
twist near the flux-rope center. The non-P.E. events also show
slightly higher twist on average around 4-5 turns/AU than most
P.E. events of about 2-3 turns/AU. Quantitatively, the average
(median) value of all P.E. associated events is 3.3 (2.8), and
that for all non-P.E. events is 5.3 (4.2), respectively. If we
exclude the point of the maximum standard deviation for each
group, the above values become 3.4 (2.4) and 4.1 (4.2),
respectively. {Note that the events of uncertain association with
P.E. are excluded from these statistics. }

\section{MEASURING PROPERTIES ON THE SUN}\label{sec:flux}
As in \citet{Qiu2007}, we here measure the reconnection flux in these
events from flare ribbon evolution observed in ultraviolet
wavelengths by the Transition Region And Corona Explorer
\citep[TRACE;][]{Handy1999} or Atmosphere Imaging Assembly
\citep[AIA;][]{Lemen2012} or optical H$\alpha$ images from the Big Bear
Solar Observatory (BBSO), combined with magnetograms obtained by
the Michelson Doppler Imager \citep[MDI;][]{Scherrer1995} or
Helioseismic and Magnetic Imager \citep[HMI;][]{Schou2012} onboard
the Solar Dynamics Observatory (SDO). Although flare ribbons form
in the upper chromosphere or transition region and the
longitudinal magnetogram is obtained in the photosphere, our
experiments have shown that using the magnetic field extrapolated to
the chromosphere changes the measured total reconnection flux by
up to 20\%. In this paper, we do not extrapolate the magnetic
field, but display the reconnection flux measured using
photospheric magnetograms, which we call $\Phi_r$ in the following
tables and text. $\Phi_r$ is measured in both positive and
negative magnetic fields, and the mean of the two is taken as the
total reconnection flux. Measurement uncertainties were
comprehensively discussed in \citet{Longcope2007b, Qiu2007,
Qiu2010}. The uncertainty mainly stems from thresholding for
flaring pixels and the difference between the fluxes measured in
positive and negative fields, which can be up to 30\%.
In the table, both the reconnection flux $\Phi_r$ and
measurement uncertainty are listed.

Apart from the reconnection flux, we also estimate the sign of
helicity of reconnection-formed flux ropes by examining the shear
of flare ribbons or post-flare loops with respect to the magnetic
polarity inversion line (PIL). Figure~\ref{fighelsign}
demonstrates how this is estimated using the example of the C2.9
two-ribbon flare that occurred at disk center on 2011 September
13. The left panel shows post-flare loops in EUV 171~\AA\ observed
by AIA, superimposed on contours of the longitudinal magnetic
field observed by HMI. The active region hosting the flare is
dominated by a bipolar configuration. In the figure, the red
(blue) contours denote positive (negative) magnetic fields with
contour levels at $\pm$ 100, 200, 400, 800~G, and the orange
dashed line roughly outlines the magnetic PIL, which is
approximated by a straight line in this case. The two flare
ribbons are parallel to the PIL, but the time sequence of ribbon
evolution as well as the orientation of the post-flare loops
reveal that post-flare loops are sheared with respect to the PIL.
The orange arrow in the figure indicates the direction of magnetic
field at the loop top along the observed post-flare loops. If the
flux rope is formed by reconnection, the shear of the post-flare
loops allows us to judge the sign of the twist of the flux rope.
We approximate this flare morphology by a 2.5d geometry, with the
translational direction along the PIL; the shear configuration
indicates the presence of the magnetic guide field, or the axial
component of the flux rope field, along this same direction
pointing from positive to negative polarity. The right panel of
the figure is a sketch of the cross-section of the assumed flux
rope structure and post-flare loops beneath it, viewed along the
PIL from the { southwest}. The magnetic configuration suggests
that the flux rope is left-handed in this event. With this method,
we estimate the sign of flux rope helicity as left-handed (L) or
right-handed (R) for the majority of events, as listed in Table 2.
Note that in some events, there is no evident shear of ribbons or
post-flare loops, or the magnetic field of the flare region is too
complex to be approximated by a bipolar structure, so the sign of
the flux rope helicity is undetermined (marked as ``U" in the
table).

Finally, the flux rope structure, if formed by reconnection, is
related to the sequence of magnetic reconnection
\citep{Longcope2007b,Qiu2009} which dictates change of
connectivity and therefore exchange of helicity between different
magnetic structures. Without applying a detailed topology
analysis, we only report the simple morphology sequence of flare
ribbons, by recognizing the apparent spreading patterns of flare
ribbons. In most eruptive two-ribbon flares, flare ribbons are
brightened simultaneously at multiple locations along the PIL, and
the two ribbons exhibit expansion perpendicular to and away from
the magnetic polarity inversion line, much resembling the 2d
standard CSHKP configuration. A good number of two-ribbon flares
are also observed to start brightening at a certain location on
the ribbon, and brightening systematically spreads along the PIL
to form the full length before expanding perpendicularly to the
PIL. Qualitatively, the first type may be interpreted as
reconnection associated with flux rope eruption that disturbs the
global magnetic field and triggers reconnection at multiple places
along the macroscopic current sheet, and the immediately ensuing
perpendicular expansion of the ribbon reflects reconnection of
overlying arcades, as depicted by \citet{Moore2001}. The initial
parallel expansion of flare ribbons along the PIL, on the other
hand, clearly violates the 2d configuration, although the
organized pattern of ribbon spreading likely implies the presence
of a macroscopic current sheet in the corona. The parallel
spreading of the ribbon may indicate sequential reconnection
between adjacent sheared arcades \citep{Longcope2007b}, in favor
of injecting a large amount of twist into the flux rope.
Furthermore, whether reconnection starts simultaneously at
multiple locations along the PIL or takes place locally and then
spreads in an organized manner may help diagnose the initial
triggering mechanism. For example, \citet{Shepherd2012} have shown
that spreading of reconnection along the PIL is likely caused by
dynamics in the current sheet. In this spirit, we also report the
pattern of morphological evolution of flare ribbons in this paper.
In Table~\ref{table2}, we use $\perp$ to indicate perpendicular
expansion of the ribbon, and $||$ to denote the presence of
parallel spreading, and ``U" refers to flare evolution not
exhibiting organized patterns most likely due to the complex
magnetic structure of the flare. It is also noted that parallel
spreading often occurs at the start of the flare; therefore, flare
observations with a low cadence might not capture such evolution
pattern during the initial phase.

\section{COMPARISON OF FLUX ROPE PROPERTIES WITH SOLAR SOURCES}
\subsection{Magnetic Flux Budget}
% flux-flux comparison
As discussed earlier, the sign of helicity between the flux ropes
embedded within ICMEs and their solar source regions compares very
well, where the topology of the erupting field and subsequently
the helicity sign of the corresponding flux-rope structure were
inferred based on Figure~\ref{fighelsign}. They agree to a large
extent (see Table~\ref{table2}, the 4th column). There is only one
mismatch, event \#12, among the 14 events with both signs
identified. {As discussed in Section 2, for this event, it is very
difficult to establish a one-to-one association between the MC and
the solar source due to a chain of flare and filament eruptions
throughout the day. The mismatch may suggest that the C3.2 flare
might not be the solar source of the MC flux rope. However,
\citet{Torok2011} modeled the three filaments as flux ropes, all
of them also carrying a left-handed twist based on observations.
Therefore, it is most likely that interactions between different
CMEs from different regions on the Sun make it difficult to
determine the helicity of the flux rope from only local magnetic
field configurations \citep[e.g.,][]{Schrijver2011}.} In addition
to such a successful comparison, we compare the magnetic flux
content of the flux ropes with that of their solar progenitors,
namely, the magnetic reconnection flux associated with preceding
flaring activity, following the original study of \citet{Qiu2007}.
We augment the original list of 9 events and show the magnetic
flux comparison among $\Phi_p$, $\Phi_t$, and the corresponding
flare-associated magnetic reconnection flux $\Phi_r$ in
Table~\ref{table2}. Note that for the previously presented events
\#1-9, the results here were further refined and improved.
Especially for event \#5, the maximum axial field and flux were
updated from \citet{Qiu2007} in the present study.

 Figure\,\ref{figflux} shows a comparison of all events
 in Table~\ref{table2} that have pairs of ($\Phi_p$, $\Phi_t)$ and
 ($\Phi_p$, $\Phi_r)$ available. It also includes one additional
 event from \citet{2009JGRAM} where a detailed study of 22 May
 2007 event was carried out and relevant quantities were obtained by the GS method.
The results generally indicate that $\Phi_p\approx 3\Phi_t$ and
$\Phi_p\lesssim\Phi_r$ for an axial effective length $L=1$ AU with
uncertainty range $L\in [0.5, 2]$ AU, which confirms the previous
results \citep{Qiu2007} although the one to one correlation
between $\Phi_p$ and $\Phi_r$ deteriorates for the enlarged
sample. One caveat associated with the few low points in the right
panel is that the poloidal MC flux was significantly
underestimated due to selection of a rather short interval for the
GS reconstruction (a few hours as opposed to normally tens of
hours) in some cases. For example, the two squares in
Figure~\ref{figflux} (right panel) of the lowest $\Phi_p$ values
correspond to events  \#15 and 16 (open symbols), respectively.
The durations of the GS reconstruction intervals are  5.2 and 2.0
hours, which yield small-size flux ropes such that each just
corresponds to a small portion of the entire ICME complex.
Therefore these small-duration GS reconstruction results would
likely lead to a significant underestimate of the ICME/MC flux.
Nonetheless, a good number of points are clustered around the
dashed line, indicating a good correlation between $\Phi_p$ and
$\Phi_r$, taking into account the associated uncertainties. {There
are generally no clear distinctions between P.E. and non-P.E.
events, except for the P.E. event \#11 (the lower left filled
square above the dashed line in the right panel of
Figure~\ref{figflux}) and \#14 (not shown, but see Table~2) of
significantly greater poloidal flux than the corresponding
reconnection flux. We will describe and discuss the former event
in Section~\ref{case} in much more detail. For event \#14,
although the association with P.E. is uncertain, the existence of
excessive $\Phi_p$ with respect to $\Phi_r$ from our analysis
implies plausible contribution from pre-existing structure such as
a filament prior to eruption.} Additionally, a major
flare-dominant event \#18, that has both flux content well
determined and falls along the one-to-one line (open square) in
Figure~\ref{figflux} (right panel), will also be presented as a
detailed case study in Section~\ref{case}.
%Whether or not that indicates the
%significant contribution from the pre-existing flux rope
%(prominence) is worth pursuing.

%\input{jiong_case2}

\subsection{Case Studies}\label{case}
The scatter plot in Figure~\ref{figflux} using extended samples in
general agrees with the previous results by \citet{Qiu2007}. For
these events, the reconnection flux measured in two-ribbon flares is
comparable with the MC poloidal flux per AU, and statistically
there is no evident bimodal distribution distinguishing events
associated with filament eruption from those without filament
eruption. For these events, the mean ratio of poloidal flux to
toroidal flux approaches 3. If we assume a uniform twist
distribution in the flux rope, this ratio yields the mean twist of
the flux rope to be about 3 turns/AU, which is above the
theoretical Kink-instability threshold. The simple estimate would
tend to suggest that reconnection would contribute significantly
to the amount of twist in these flux ropes, even if these flux
ropes were pre-existing with a smaller amount of pre-existing
twist to start with.

Nevertheless, the plot also reveals a few outliers deviating from
the general pattern of flux-flux comparison. The MC associated
with a B-class flare on 2010 May 23 (event \#11) carries a
significantly larger poloidal flux, which is about 3 times the
reconnection flux measured in the minor flare, indicating that a
large amount of poloidal flux cannot be contributed by
reconnection. On the other hand, this MC also possesses a
relatively large toroidal flux, and as a result is less twisted
than the majority: the mean ratio of poloidal flux (per AU) to
toroidal flux is 2. Furthermore, analysis of the structure of the
MC shows a rather flat twist distribution from the core of the
flux rope outward, with $ \langle\tau\rangle \approx 2$ turns per
AU and a standard deviation about 20\%.

In contrast to this event, which is likely a case of a dominant
pre-existing flux rope, the event that occurred on 2011 September 13-17
(event \#18) well fits the scenario that reconnection may
dominantly contribute to the poloidal flux of the MC. In this
event, $\Phi_p \approx \Phi_r$, and $\Phi_p \approx 3.6 \Phi_t$.
Furthermore, the MC is shown to be highly twisted at the core,
with a twist value about 5 turns/AU and higher, which decreases
outward to about 3 turns/AU (see Figure~\ref{fig0914_2}). In this
case, it may be reasoned that a flux rope with such a large amount
of twist would be subject to Kink instability, and therefore
cannot pre-exist stably prior to eruption. The event is a case in
favor of the scenario that the highly twisted flux rope is largely
formed by reconnection during the eruption.

MCs associated with these two events are well measured by Wind/ACE
at 1~AU with little ambiguity in GS reconstruction results,
showing typical large-scale flux rope structure of similar sizes
and magnetic field strength. The flares and CMEs associated with
the MCs are also very well observed by AIA and STEREO,
respectively. Therefore, we choose these two events for detailed
analysis of their solar progenitors, namely flares and CMEs, to
understand whether there is a meaningful difference in the
solar surface signatures between the two events that have quite
different MC structures, especially in terms of field-line twist
distributions.

\subsubsection{Flare/CME/MC Event in 2011 September 13-17}
Figure~\ref{fig0914_1} gives a panorama view of the C2.9
two-ribbon flare observed by SDO and its associated CME observed
by STEREO. The flare occurs in a nearly bipolar magnetic
configuration (Figure~\ref{fig0914_1}c), with one flare ribbon
first brightened at the northwest end and then spreading along its
own length of 50 Mm over the course of less than an hour
(Figure~\ref{fig0914_1}a and b). The apparent uni-directional
parallel spreading at a mean speed of 16 km s$^{-1}$ is much
slower than characteristic Alfv\'{e}n speed, so the apparent
motion pattern is likely governed by spreading of reconnection
sites due to current drifting along the overlying macroscopic
current sheet in the corona \citep{Shepherd2012}. Analysis of the
flare ribbon evolution yields measurements of time-dependent
reconnection flux, plotted in Figure~\ref{fig0914_2}, showing that
the reconnection flux amounts to 6$\times 10^{20}$ Mx within an hour
from the flare onset, with a peak reconnection rate of 2$\times
10^{17}$ Mx s$^{-1}$ at 23 UT on 2011 September 13. The
uncertainty in reconnection flux shown in the plot mainly reflects
the imbalance between the fluxes measured in positive and negative
magnetic fields. The sequential reconnection and formation of
flare loops are also manifested in the sequence of post-flare
loops observed in a few EUV bands by AIA. The second row of images
in Figure~\ref{fig0914_1} shows the first appearance of post-flare
loops, observed in EUV 171\AA\, in the northwest, which then
``spread" downward along the PIL. These loops are anchored at the
ribbons that had brightened in UV emission 20 minutes earlier.

It is also noted that EUV dimming, or reduced EUV emission in the
171\AA\ hand, is observed prior to the appearance of post-flare
loops. To compare the timing of dimming with the flare/reconnection
process, in Figure~\ref{fig0914_2}, we plot the time profile of
the inverted total EUV flux in the flaring active region together
with the reconnection flux. In this plot, the rise of the dimming
curve indicates decreased total EUV flux at the 171\AA\ band in
the active region, and the decay of the dimming curve at 0UT of
2011 September 14 indicates enhanced EUV emission in post-flare
loops formed by reconnection. It is seen that the dimming curve
rises on the same timescale as the reconnection flux. A careful
examination of high-cadence (10~s) high-resolution
($\sim$1\arcsec) imaging observations in Figure~\ref{fig0914_1}
suggests that EUV dimming is primarily caused by disruption and
disappearance of a few sets of pre-flare active region loops at
the time of reconnection, as evident in a comparison between
panel (d) and panel (e) and the difference image of these two
images in panel (f). The morphology of dimming well tracks the
shape of the pre-flare coronal loops from their feet to the top. Some
of these loops implosively disappeared, most likely due to
re-organization of pre-existing magnetic structures by
reconnection. These disrupted pre-flare loops marked in the figure
also appear to be more sheared than the post-flare loops that formed
underneath twenty minutes later. As the dimming morphology largely
tracks the shape of the loops, we cannot unambiguously interpret
dimming entirely as being produced by evacuation of coronal
plasmas at the locations where the flux rope is rooted and ejected
\citep{Webb2000}.

By careful scrutiny, we can identify three locations of dimming at
the feet of disappearing loops. These three locations are marked
as``D1", ``D2", and ``D3", respectively, in
Figure~\ref{fig0914_1}f. D1 is located in a sunspot of negative
magnetic fields, and the other two reside in plages of positive
magnetic field. D1 and D3 exhibit dimming starting at the onset of
reconnection at 22 UT on 2011 September 13, and peaking two hours
later. At D2, dimming starts half an hour later at 22:30~UT but
peaks earlier at 23~UT. The dimming in all places then persists at
the same flux levels until 4UT next day when the flux starts to
recover very slowly. Some of these locations may be where the flux
rope is rooted, and the magnetic flux summed in these regions provides
an estimate of the toroidal flux in the ejected flux rope. The
negative flux estimated in the strong magnetic field of the
sunspot carries a lot greater uncertainty than the flux measured
in the weak positive fields in the plage regions, because of the
large amount of magnetic flux in the regions of projection of
disrupted magnetic loops which are difficult to distinguish from
the feet. Therefore, we only measure magnetic flux in D2 and D3,
which turns out to be $\Phi_{D2} = 1.9\times 10^{20}$ Mx and
$\Phi_{D3} = 1.5\times 10^{20}$ Mx, respectively. These numbers
are close to the toroidal flux measured in MC, $\Phi_t = 2.4\times
10^{20}$ Mx, although it is hard to judge which of the two regions
is more likely the foot of the finally ejected flux rope.

The CME associated with this flare is observed by all three
instruments, EUVI, COR1, and COR2, onboard STEREO. STEREO-A allows
a better view of the CME, as shown in the third row of the figure.
In the EUV 195\AA\ images by EUVI, a coronal structure hanging at
the height of 1.25 solar radii is vaguely visible prior to the
onset of the flare at about 21:45 UT on 2011 September 13, and
very slowly rises at almost a constant speed. The structure and
its movement become evident when reconnection on the disk takes
place at 22~UT, and the CME is subsequently observed in the COR1
and COR2 field of views (FOVs). The CME exhibits a circular front followed by a
core structure beneath. To track its movement, we construct a
time-distance plot along a straight slit connecting the solar
center with the top of the rather circular CME structure. These
plots constructed using base difference images by EUVI, COR1, and
then COR2 are illustrated at the bottom row of the figure, which
clearly outline the CME core in all three types of images as well
as the front in COR1 and COR2 images. We then made an automated
routine to measure the height of the CME core by following the
maximum intensity in the core structure, and the half width of the
core structure is taken as the measurement uncertainty. The
measurement is shown in the blue curve in Figure~\ref{fig0914_2},
against the reconnection flux plot. The CME rises slowly in the
first 40 minutes, and then speeds up at a height of 1.5 solar
radii.

To derive its velocity, we make a piece-wise linear fit to the
measured heights versus times for data points to up to 5 solar
radii; beyond that distance, the CME structure spreads out giving
large uncertainties in determining the centroid of CME mass.
Uncertainties in the velocity measurements are simply standard
deviations of the linear fit to each piece. As shown in the bottom
left panel of the figure, the CME reaches the maximum speed close
to 300 km s$^{-1}$ at 23 UT at around 2 solar radii. The
acceleration is obtained by taking time derivatives of the
velocity, and error bars are derived from error propagation. It
appears that peak acceleration, of order 80 m s$^{-2}$ occurs when
the reconnection flux rises most rapidly at around 23 UT of 2011
September 13, which is consistent with some previous results,
though some of these earlier measurements have used lower-cadence
CME data
\citep{Zhang2001,Qiu2004,Patsourakos2010,Patsourakos2013,Cheng2014}.
Reconnection nearly stops after midnight, when the flux rope is at
2.5 solar radii. In addition, the height of the CME front is
measured in the same way and plotted in violet in the top left
panel. It is probably a compression shock front driven by the CME.
Below 2.5 solar radii, the CME velocity reaches over 300 km
s$^{-1}$, fast enough to drive a shock front.

It is evident in these plots that CME acceleration is coincident
with the progress of magnetic reconnection. If reconnection
injects magnetic flux into the CME structure, and if the CME is
assumed to undergo a self-similar expansion, in which case, the
size of the CME flux rope $R_{fr}$ grows proportionally with the
height of the CME $H_{fr}$, then we can estimate the rate of flux
injection as a function of the size of the infant flux rope when
it is close to the Sun, e.g., $H_{fr} \le 2 R_{\sun}$. The upper right
panel of Figure~\ref{fig0914_2} shows the reconnection flux
($\Phi_r$; red) and reconnection rate (black) against the height
($H_{fr}$) of the CME core, and the blue curve shows the rate of
the flux injection defined by $\psi_{fr} = d\Phi_{r}/dH_{fr}$. The
injection rate rises rapidly with CME height and peaks at a
height of $H_{fr} \approx 2 R_{\sun}$ with $8\times 10^{20}$ Mx
R$_{\sun}^{-1}$. As reconnection slows down and eventually stops,
the flux injection ceases. The field-line twist distribution
within the flux rope at 1 AU as depicted in the lower right panel
exhibits a clear and largely monotonic decline from the center
to about 1/3 way through the interval, then remaining flat toward
the boundary.

%   ; ok here piece-wise linear fit, and therefore piece-wise acceleration.
%   ; 60 km/s from 21:45 - 22:00 ste(1:9)
%   ; 80km/s from 22-22:30 (ste(10:*)), mean acceleration of 17 m/s^2 at middle time 21:52 UT
%   ; 150 km/s from 22:50 - 23UT (st1(6:9)), mean accel. of 30 m/s^2 at middle time of 22:35
%   ; 245 km/s from 23 to 23:42 UT (st1(10:*)), mean accel. of 61 m/s^2 at middle time of 23:08 UT.
%   ; 228 km/s from 23:34 to 24:34 UT (st2(indgen(6)*6)), so zero acceleration
%   ; 342 km/s from 23:46 to 27:22 UT (st2(indgen(6)+14)*6), mean accel is 22m/s^2 at 24:50 UT.
%of 60 km s$^{-1}$ till 22 UT. Between 22 and 22:30 UT, the CME is mainly visible in EUVI FOV rising with
%a speed of 80 km s$^{-1}$. The rope is clearly seen in COR1 FOV after 22:50 UT, and rises
%at the speed of 150 km s$^{-1}$ for the next few minutes, and then from 23 to 23:40UT, another
%linear fit gives the speed of 245 km s$^{-1}$. Therefore, fast acceleration occurs just
%around 23 UT with peak acceleration of 170 m s$^{-2}$. OK, ONLY DO A CLOSE-TO-SUN PLOT OF PIECE WISE
%LINEAR FIT TO GET VELOCITY AND ACCELERATION FOR A FEW POINTS.

\subsubsection{Flare/CME/MC Event in 2010 May 23 - 28}
In the same way, we present the images and plots for the flare/CME
event on 2010 May 23. The top panels of Figure~\ref{fig0528_1}
show that the two-ribbon flare evolution, in contrast to the other
event, nearly follows the 2d CSHKP model with ribbons brightening
at multiple locations along the PIL, and then expanding vertically
outward in a nearly 2d manner. The reconnection flux measured in
this event is plotted in Figure~\ref{fig0528_2}. For this event,
the total reconnection flux amounts to 2.7$\times 10^{20}$ Mx,
which is only one third of the measured poloidal flux in the MC
observed 5 days later. Figures~\ref{fig0528_all} and
\ref{fig0528_3D} show the corresponding GS reconstruction results
of the MC flux rope from Wind spacecraft data.

EUV dimming is also observed. Unlike the other event, the dimming
plot in the top left panel does not track the reconnection flux
plot very well; it rises more gradually than reconnection flux. At
some locations, dimming appears to be removal of pre-flare coronal
loops, as in the case of the other flare. But the dimming
morphology in this event also exhibits some differences. It is
seen in EUV 171 images that dimming also occurs along the
locations of flare ribbons before they are brightened immediately
afterwards. This morphology evolution much resembles the scenario
depicted by \citet{Forbes2000} and \citet{Moore2001}, that the
erupting flux rope stretches overlying coronal field lines, which
then close down by reconnection. There is also a patch located in
an EUV moss region next to the ribbon (indicated by the arrow in
panels (e) and (f) in Figure~\ref{fig0528_1}), which does not
appear to be parts of high-lying coronal loops. Dimming takes place
in the patch by removal of the moss structure and spreads outward
in a way very similar to the event reported by \citet{Webb2000},
making it a viable candidate for a foot of the erupting flux rope.
The patch is located in negative magnetic field, and magnetic flux
measured in this dimming patch amounts to 3.0$\times 10^{20}$ Mx,
similar to the MC toroidal flux 3.8$\times 10^{20}$ Mx. It is,
though, not clear from observations where the other foot
of the erupting flux rope is located.

The CME is prominent in the views of COR1 and COR2 onboard
STEREO-B. In the STEREO EUVI images, the erupting structure itself
is invisible; however, abrupt dimming was observed around 16:30 UT
(panel (g) in Figure~\ref{fig0528_1}) suggesting occurrence of
eruption that expels nearby plasmas. Around this time, the CME
front can be observed in the COR1 images. The CME core itself is
first seen in the COR1 image at 17~UT. The time-distance plot
along a slit connecting solar center and the top of the CME
structure is displayed at the bottom panel of
Figure~\ref{fig0528_1}, from which we measure the height of the
CME core as well as its front shown in the top left panel of
Figure~\ref{fig0528_2}. The bottom panel shows the CME velocity
derived from a piece-wise linear fit and the CME acceleration
obtained from time derivatives of the velocity. The CME evolution
is very similar to the other event on 2011 September 13-14: {both
events experience a short period of fast acceleration, which peaks
around the time reconnection also peaks.} Both events reach a
maximum velocity of 300 km s$^{-1}$, and both arrive at 10 solar
radii six hours after onset of eruption.

This event has a much smaller reconnection flux than the
other one, although they exhibit very similar CME evolution.
Suppose that this is the same amount of flux injected into the
erupting flux rope, then the flux injection rate per solar radii
of the CME height is smaller by more than half an order of
magnitude. The twist distribution for this event remains fairly
flat, at about 2 turns/AU, throughout the flux-rope structure as
shown in Figure~\ref{fig0528_2} (lower right panel). The rapid
increase of the green curve toward the outer boundary is due to
increased errors in this estimate (see Appendix).
%It is also noted
%that reconnection, and therefore the assumed flux injection, takes place more gradually in
%this event, which spans over 2.5 hrs and from 1.6 to 3 solar radii. In comparison, in the other
%event, flux injection proceeds for about 1.5 hrs when the CME rises from 1.3 to 3 solar radii.

\subsection{How Reconnection Affects CMEs}
Joint observations by SDO and STEREO from different view points
and with unprecedented tempo-spatial resolution allow us to track
kinematic evolution of CMEs in their infancy from as low as
250\arcsec\ above the surface, and at the same time reliably
measure properties of reconnection beneath the CME flux rope. From
comparison of reconnection properties and CME properties in the
two well-observed events, it is evident that prominent
acceleration of the CMEs of order 100-200 m s$^{-2}$ takes place
during the first 1-2 hrs when reconnection proceeds rapidly; in
this stage, the CME flux rope speeds up from a few tens of km
s$^{-1}$ to a few hundred km s$^{-1}$ and from the height close to
the Sun (1.3-1.6 solar radii) to about 3 solar radii. These
results confirm, with observations of much better quality, the
suggestion from previous flare-CME observations that CME
acceleration and magnetic reconnection manifested in flares appear
to be temporally correlated
{\citep{Zhang2001,Qiu2004,Qiu2005,Jing2005,Patsourakos2010,Temmer2010,Patsourakos2013,Cheng2014}}.
Physically, it is not difficult to see why this should happen:
reconnection changes the magnetic configuration, which inevitably
changes the magnetic forces acting on the flux rope. In the
specific cases discussed in this paper, it appears that such
changes would result in an overall expulsion force on the flux
rope. Just by contrasting these two events, it also seems that the
effect of reconnection on the kinematic evolution of CMEs is not
qualitatively different in a pre-existing flux rope and an in-situ
formed flux rope. The coincident onset of fast reconnection and
major acceleration is recently revealed in advanced
magnetohydrodynamic (MHD) simulations that use refined and
adaptive grids to resolve the role of core reconnection in CME
acceleration \citep{Karpen2012}.

A more interesting and indeed critical question concerns whether
and how reconnection also changes the structure of the flux rope
itself. Most theoretical as well as numerical models of CME
eruption would envision, at least qualitatively, injection of
magnetic flux into the CME flux rope. If this happens, then
properties of reconnection in a time sequence would be responsible
for the structure of the infant flux rope from its core outward.

We can discuss three different models, all in a 2.5d scheme, of
reconnection and its effect on flux rope structure on the
Sun. In a standard strict 2d CSHKP model, eruption of a certain
kind of pre-existing flux rope such as embodied in a filament
pulls the overlying arcade, which reconnects with itself below
the ejecting flux rope \citep[e.g., the cartoon model
by][]{Moore2001}. This process produces a bubble of field lines around
the axis, or adds poloidal flux around a constant pre-existing
toroidal flux. Recent 3d numerical simulations have shown
characteristics of such bubble-field lines added to the flux rope
\citep{Aulanier2012} in the later stage of reconnection.
Noteworthily, \citet{Aulanier2012} also illustrates the earlier
stage of flux rope formation, showing a less twisted flux rope at
the start, with more twisted flux added to it as reconnection
proceeds between overlying coronal fields. The simulation is used
to interpret the observed apparent shear motion of flare ribbons.
We note that in these scenarios, the infant flux rope
would therefore have low twist at its axis, followed by higher
twist outward. However, we do not find many examples of this kind
in the analyzed MCs in this sample. Most MCs exhibit either a flat
twist distribution or twist decreasing from the core outwards.

Another type of reconnection proposed by
\citet{Ballegooijen1989} is that reconnection takes a few steps to
first form the flux rope axis, which is a long sheared loop along
the polarity inversion line; in the following steps, reconnection
takes place between a pair of sheared arcades both above the
primary axis, and results in a loop twisting around the primary
axis. In this process, different from the 2d model in which
reconnected field lines are detached from the solar surface, {\it
toroidal} flux is injected into the flux rope by the amount of
flux carried in one sheared arcade prior to reconnection, and {\it
poloidal} flux is also injected, and the amount of added twist is
roughly 1.5 turns, i.e., the newly added field line makes one and
half turns from end to end. If this process continues with more
and more pairs of overlying field lines reconnecting with each other,
but only once, then the net consequence is that the flux rope is
formed with increasing toroidal flux and a flat twist of 1.5
turns. We suggest that 2010 May 23 - 28 event may be described by
this pattern, with the entire process of flux rope formation
taking place in at least two different stages, the first stage
being formation of the flux rope filament prior to the flare, and
the second stage during the B-class flare, that injects toroidal
as well as poloidal flux into the rope, but with a constant twist
distribution of about 1.5. Evidence of such a process includes:
the short dimming along later brightened flare ribbons indicating
stretching of a set of arcade field lines prior to reconnection by
eruption, immediately followed by simultaneous brightening of two
flare ribbons at multiple locations along the PIL, and then
dominant apparent ribbon motion perpendicular to the PIL
suggesting progressive reconnection by higher loops. This event
also exhibits a significant dimming patch next to the flare
ribbons, which is not brightened later on suggesting that no
reconnection takes place at this location. Morphology of this
dimming patch is very similar to that of \citet{Webb2000}. We
suspect this is one foot of the primary axis of the pre-existing
flux rope. Such a dimming morphology is not observed in the other
flare discussed below.

In the third scenario as demonstrated by \citet{Longcope2007a},
the first step reconnection takes place between a pair of sheared
arcades to form a flux rope with one turn and an underlying
post-flare loop; in the following steps, the flux rope continues
to reconnect with adjacent sheared arcades sequentially. Each step
injects more twist into the rope whereas maintaining the toroidal
flux, which is the amount of the flux from the first set of
reconnecting arcades. We propose that the event on 2011 September
13 - 17 exhibits a few observational signatures indicative of this
process though only qualitatively and possibly mixed with other
processes. First it is evident that some pre-flare sheared loops
disappeared at the onset of the flare, producing dimming flux that
evolves on the same timescale of reconnection flux; the post-flare
loops formed later on are beneath these pre-flare arcades and are
less sheared. It is likely that dimming, or disappearance of
pre-flare loops, is largely caused by reconnection of pre-existing
sheared arcades. Second, reconnection as inferred from evolution
of both the flare ribbons and post-flare loops exhibits a very
regular sequence starting from one end of the ribbon proceeding to
the other end, much as predicted in the sequential reconnection
model. This reconnection sequence along the PIL is then followed
by ribbon spreading perpendicular to PIL, but the second stage of
perpendicular spreading is insignificant compared with the first
stage of dominant parallel spreading. We suggest that the first
stage produces high twist in the inner part of the flux rope,
whereas the second stage plays a role similar to the second
scenario that would add toroidal flux and a flat twist in the
outer part of the flux rope by reconnection between adjacent
overlying arcades. In particular, the dominant early-stage
sequential reconnection along the PIL at a speed of 10-20 km
s$^{-1}$ is hard to be explained by an erupting pre-existing flux
rope stretching field lines and triggering reconnection, in which
case, the coronal field would be violently disturbed at multiple
locations and therefore reconnection would take place in multiple
locations without a prescribed order along the PIL. In other
words, such an observed reconnection sequence would be in favor of
reconnection governed locally such as by resistive instabilities
or current sheet dynamics than reconnection driven by MHD
instabilities \citep{Karpen2012,Shepherd2012}.

We recognize that there remain a couple of observational details
pending explanation with this scenario, one being posed by the
STEREO-EUVI observation that an overlying coronal structure is
present about 15 minutes before the observed onset of the
sequential reconnection, and its evolution later on appears to be
consistent with the CME core (the flux rope) identified in COR1
and COR2 images. It is not clear what is the relation between this
structure and the flux rope being formed by sequential
reconnection. It is possible that weak reconnection and formation
of the flux rope already starts before 22 UT but with very weak
signatures on the disk. Another detail concerns the rather long
timescale of reconnection in this event, which proceeds for 60
minutes, with the fast reconnection and organized pattern of
spreading lasting for 40 minutes from 22:50 to 23:30 UT. During
this period, the STEREO-observed CME core moves from 0.5 to 1.5
solar radii above the limb. The connection between the flux rope
and coronal reconnection would imply the presence of a long current
sheet linking the bottom of the flux rope and the top of the
post-flare arcade. Furthermore, whereas the flux rope moves
rapidly in the high corona, reconnection below the flux rope
proceeds in an organized ``zipper" pattern, which may suggest that
there is only very weak overlying coronal magnetic field. It is
possible that a pre-flare break-out type reconnection has taken
place to remove much of the magnetic flux above the core flux
rope.

\section{CONCLUSIONS AND DISCUSSION}
In conclusion, we have analyzed magnetic clouds and their solar
progenitors including flares, CMEs, and coronal dimming for an
enlarged sample containing a total of 19 events. The magnetic
structure of flux ropes is examined by the GS reconstruction
method, and is compared with the properties of flares, filaments,
and coronal dimming in the corresponding source region. We summary
our main findings as follows.

\begin{enumerate}
\item{Our systematic analysis of the magnetic field-line twist distribution within magnetic flux
ropes provides clear evidence for the invalidity of the 1D
constant-$\alpha$ force-free model of a cylindrical flux rope.
Such a model predicts increasing twist with increasing radial
distance away from the flux-rope center, approaching infinity at
the boundary where $B_z=0$. However our analysis does not
show this general trend. Instead our results are more consistent
with a non-linear force-free model. In about half of the cases,
the field line twist is constant at 1.5-3 turns per AU, as in the
Gold-Hoyle (GH) model. The other half exhibit a high
twist of $\gtrsim 5$ turns per AU near the core, which decreases
outward. There is suggestion that events associated with
filament eruptions have a lower average twist than
events not associated with filaments.}

\item{We compare the MC magnetic structure with properties of solar flares associated with the CME/MC.
It is shown that the sign of the helicty of MCs is consistent with the sign of helicity of the post-flare
coronal arcade, and the amount of twisted flux (the poloidal flux) in general agrees
with the measured amount of flux reconnected in flares. There is no statistically significant
difference between events with or without filament eruption.}

\item{We also conduct detailed case studies of two events with typical
and comparable flux-rope geometry but different twist distribution,
one with a flat twist of about 2 turns/AU from center to edge, and
the other with a high twist about 5 turns/AU near the axis, which
decreases outward. The two events are very well observed by
multiple spacecraft at multiple view points on the Sun and at 1
AU. Comparison of the MC flux and reconnection flux, as well as
the flare and dimming evolution, suggests that the first event is
probably dominated by a pre-existing low-twist flux rope surrounding
a filament, and reconnection at multiple locations along the PIL appears to
add only a small amount of flux with low twist;
whereas the second event is probably a flux rope with significant
twist injected by slow sequential reconnection along the
PIL. This case study, though limited in its scope, suggests that
the geometry of reconnection as reflected in flare morphology is
related to, and therefore may be used to diagnose, the magnetic
structure of infant flux ropes formed on the Sun. In terms of the
kinematic evolution, in both events, the onset of fast
acceleration takes place when fast reconnection starts regardless
of the geometry of reconnection.}
\end{enumerate}

The field-line twist distribution in MCs is consistent with a
constant-twist non-linear force-free model \citep{1960MNRAS.GH}.
The force-free parameter $\alpha$ changes with flux surface in
this model, although it remains constant on each distinct surface.
This implies that these flux surfaces are formed at the Sun and
are not destroyed while propagating in the interplanetary space.
It seems that finite resistivity is not playing a significant role
in merging these surfaces to give a constant-$\alpha$ relaxed
state \citep[see, e.g.,][]{HU:freidberg87,1986RvMPT}. The GS
method is applicable to a non-force free state, but in the
large-scale ICME structures we examined, the magnetic field always
dominates and the plasma pressure gradient does not play a major
role, even after including the additional contribution from $T_e$
\citep{2013SoPhH}. The indication that the flux surfaces are
probably preserved during the transit of the flux rope through
1~AU is significant and may provide validation for direct
comparison of MC properties with magnetic properties directly
measured or indirectly inferred in the Sun's corona. Ideal MHD
proves to be a good approximation in the solar wind such that the
magnetic flux and helicity are conserved, which in turn provides
the ground for our interpretations and discussions about relating
in-situ magnetic flux-rope structures as characterized by the
field-line twist (see Appendix) with their solar sources.

{ These findings pose questions on formation mechanisms of flux
ropes: what produces the high twist at the core of some flux
ropes, and what mechanism leads to formation of flat twist in some
other events. Our results hint at the scenario of reconnection
forming high twist at the core, which can be examined by
comprehensive models that investigate the pre-eruptive magnetic
field configuration, as well as the change of magnetic topology
and redistribution of magnetic helicity as a result of
reconnection.}

The biggest uncertainty in our analysis comes from the estimate of
the effective length $L$ of a cylindrical flux rope. We used a
nominal value $L=1$ AU with a wide range of uncertainty $L\in[0.5,
2]$ AU. Such a value is justified by the study of
\citet{2012SoPhK}, where a length of $1$ AU was used and yielded
consistent results in both magnetic flux and helicity
conservation for four strong flare-CME-ICME events. In a following
companion paper, we will address the length of field lines within
flux ropes employing both in-situ flux-rope modeling and
associated electron burst onset time observations
\citep{2011JGRAK,2011ApJK}. We will show that the length $L$ can
be further constrained to be between 1 and 2 AU, based on the
analysis of a handful of events in \citet{2011JGRAK}.

\acknowledgments

We thank the referee for a thorough, critical, and constructive review
that leads to the improved work. We are grateful to Dana Longcope and Eric Priest
for insightful discussion and help with improving the manuscript.
The work of JQ is supported by NSF grant ATM-0748428
and by NASA Guest Investigator Program NNX12AH50G. QH acknowledges
NSF SHINE AGS-1062050 and NASA grants NNX12AF97G and NNX12AH50G
for support. We acknowledge SDO and STEREO missions for providing high quality
observations. We thank ACE Science Center and NASA CDAWeb for
providing ACE and Wind spacecraft data.

\appendix

\section{Calculation of Magnetic Field-line Twist of a Cylindrical Magnetic Flux Rope}

For a cylindrical magnetic flux-rope model, the field lines lie on
cylindrical isosurfaces of the flux function $A$ ($A$ shells,
nested cylinders of arbitrary cross-sections) and wind along the
$z$ axis (e.g., see Figure~\ref{fig0528_3D}). The individual
field-line twist, $\tau$, in unit of turns/(unit length), where
the unit length is taken as 1 AU in this particular study, can be
approximated by various methods as described below. Since all the
field lines lying on the same isosurface have the same values of
$\tau$, the field-line twist thus becomes a single-variable
function of $A$. In the special case of axi-symmetry, it is  a
function of radius $r$ only. In some publications the twist is
defined in units of radians/(unit length) which differs from the
definition adopted here by a factor 2$\pi$.

Here we employ four different methods with the first three being
approximate and yielding an average field-line twist within each $A$
shell, whereas the forth one, dubbed the graphic method, yields
the  field-line twist for each individual field line on each $A$
shell. The first three approximations, based on magnetic flux
($\Phi_{p,t}$) and magnetic helicity ($K_r$; \citep{2010JGRAW})
 calculations, yield the following average field-line
twist
\begin{eqnarray}
\tau_H&=& \frac{K_r}{\Phi_t^2}\label{eq:tauH}\\
\tau_F&=& \frac{\Phi_p}{\Phi_t}\label{eq:tauF}\\
\tau_{dF}&=&-\frac{\mathrm d\Phi_p}{\mathrm d\Phi_t}\approx
-\frac{\Delta\Phi_p}{\Delta\Phi_t}
\label{eq:tauF2}\end{eqnarray}
 The derivations follow the works
of \citet{1984JFMB} and are briefly presented below, based on the
assumption of a constant (average) field-line twist, $\T$.

From \citet{1984JFMB}, we have (with the magnetic
helicity denoted by $H$)\begin{equation} dH=\Phi_p d\Phi_t -\Phi_t
d\Phi_p.\label{eq:dH}\end{equation} Then by definition, we have
$\T=-d\Phi_p/d\Phi_t$, equation~(\ref{eq:tauF2}) above. By
integrating both sides of equation~(\ref{eq:dH}) by parts for
$d\Phi_t$ and $d\Phi_p$, respectively, between 0  and the toroidal
flux $\Phi$ within certain boundary ($A$ shell), we obtain
(keeping $\T\equiv constant$)
\begin{eqnarray}
H&=&\Phi_p(\Phi)\Phi_t(\Phi)-\Phi_p(0)\Phi_t(0)+\T[\Phi_t^2(\Phi)-\Phi_t^2(0)],\nonumber\\
H&=&-\Phi_p(\Phi)\Phi_t(\Phi)+\Phi_p(0)\Phi_t(0)-\frac{1}{\T}[\Phi_p^2(\Phi)-\Phi_p^2(0)].\end{eqnarray}
To make the above two equations compatible with each other,  we
can apply $\Phi_t(0)=\Phi_p(\Phi)=0$ because they only vary along
distinct flux surfaces \citep[e.g.,][]{1986RvMPT} as defined by
$A$
 subject to a relative shift. Therefore
the above equations are reduced to
\begin{equation}H=\T\Phi_t^2(\Phi)=\frac{1}{\T}\Phi_p^2(0),\end{equation}
which in turn yields the estimates for $\T$ in
equations~(\ref{eq:tauH}) and (\ref{eq:tauF}).
 Clearly these twist estimates are all functions
of $A$ alone since all quantities involved in these estimates are
single-variable functions of $A$. The usual way of calculating the
twist locally by $\frac{1}{r}\frac{B_\phi}{B_z}$ is also
applicable for analytic 1D flux-rope models when the magnetic
field components are known in cylindrical coordinates.

\begin{figure}
\includegraphics[]{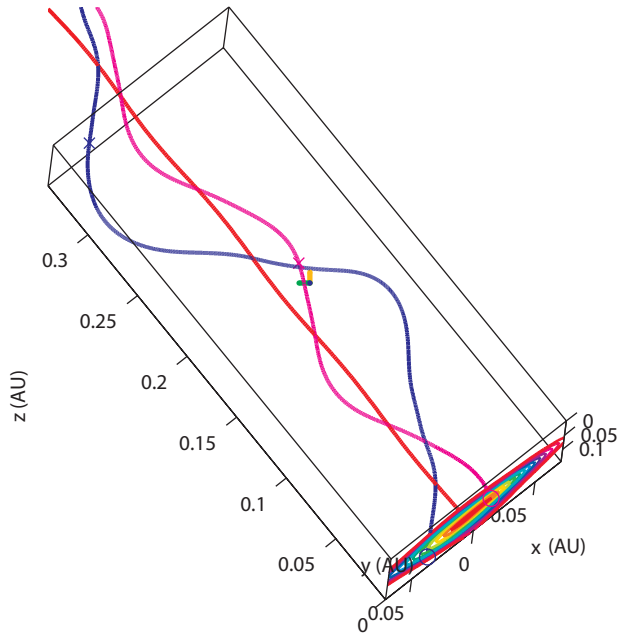}
\caption{A 3D view of the flux-rope configuration of event \#18.
The view direction is toward the Sun along the GSE-X direction,
while the GSE-Z and Y axes are pointing upward and horizontally to
the left, respectively, as denoted by the two short lines near the
center. Three field lines are shown with footpoints rooted on
$z=0$ plane where a color filled contour plot of $A(x,y)$ is
superimposed. } \label{fig0914_3D}
\end{figure}

\begin{figure}
\includegraphics[]{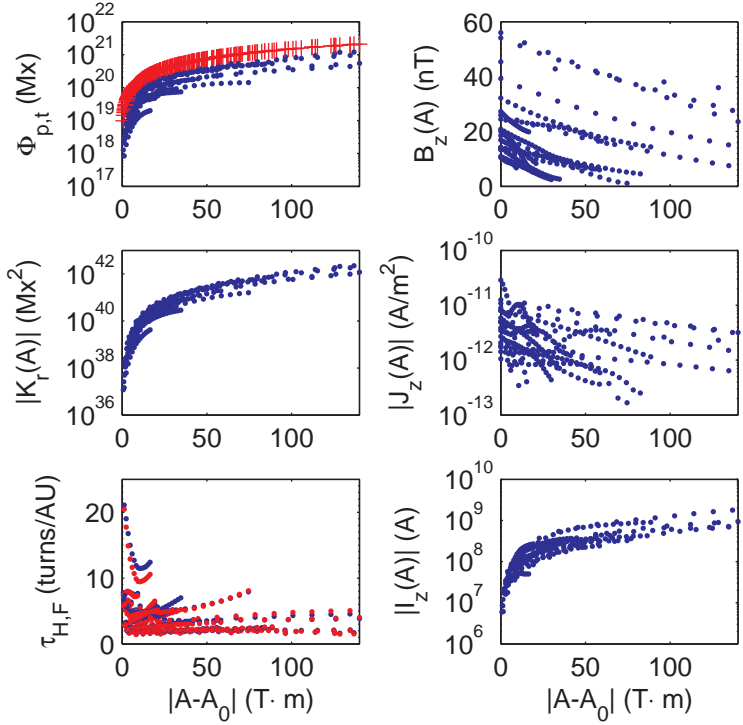}
\caption{Physical quantities (unsigned) for all events versus the
shifted flux function: (counterclock-wise from the top left panel)
the poloidal (red pluses) and toroidal magnetic flux $\Phi_{p,t}$,
the  relative magnetic helicity $K_r$, the field-line twist estimates
$\tau_H$ (red dots) and $\tau_F$ (blue dots), the axial current $I_z$,
the axial current density $J_z$, and the axial magnetic field $B_z$.}
\label{figALL}
\end{figure}

\begin{figure}
\includegraphics[width=0.5\textwidth]{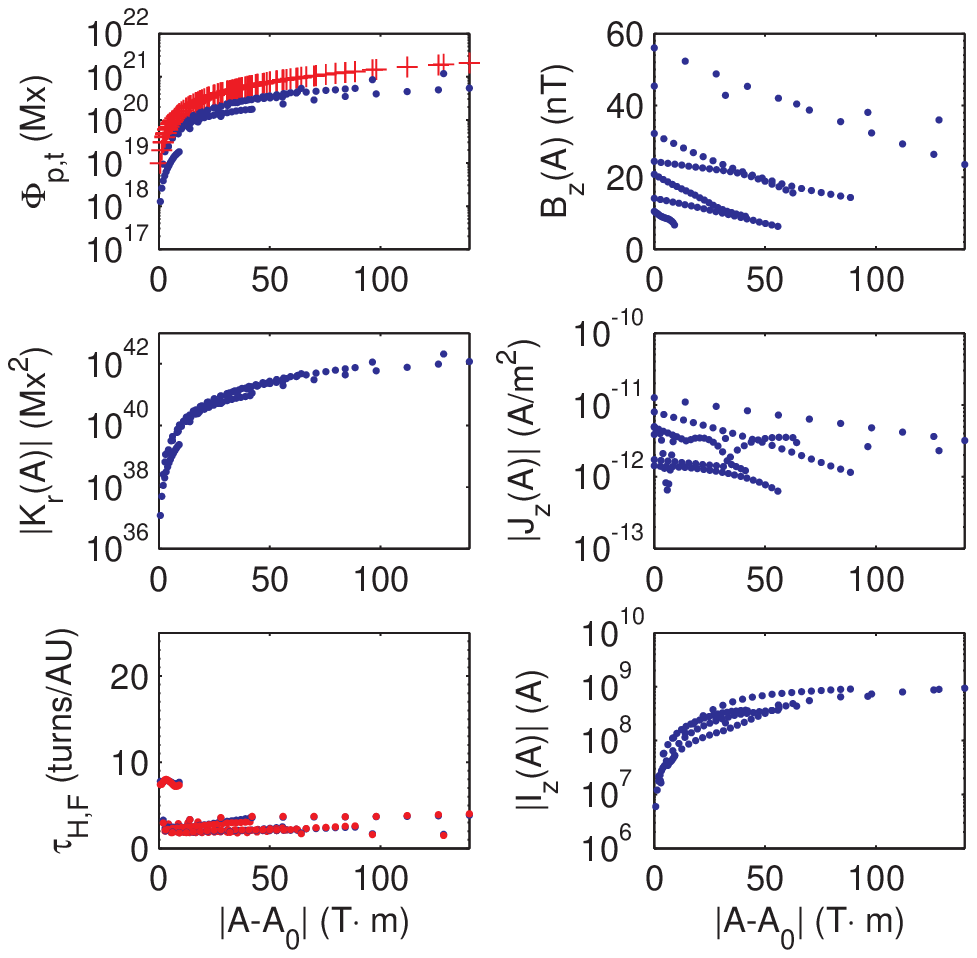}
\includegraphics[width=0.5\textwidth]{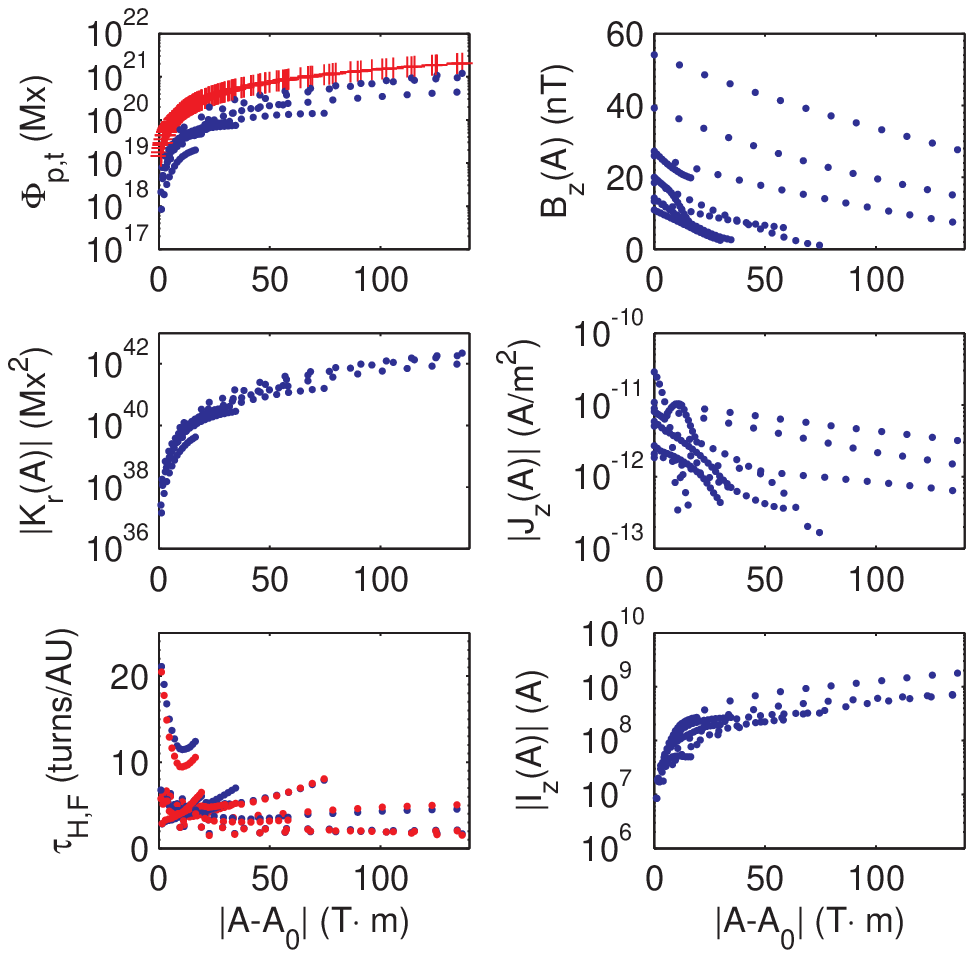}
\caption{ Distribution of various quantities for events associated
with (left 6 panels) and without (right 6 panels) prominence eruption. For each
group, the format is the same as Figure~\ref{figALL}. }
\label{figVAR}
\end{figure}

\begin{figure}
\centering
\includegraphics[]{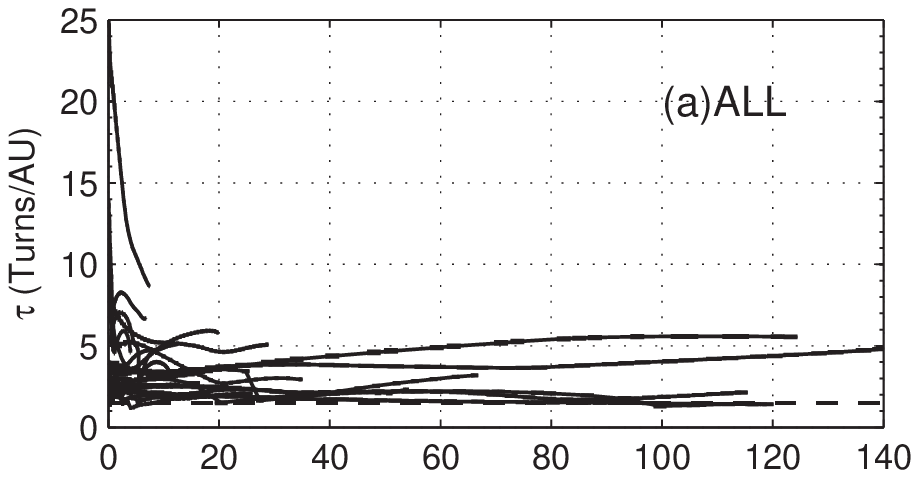}\\
\includegraphics[]{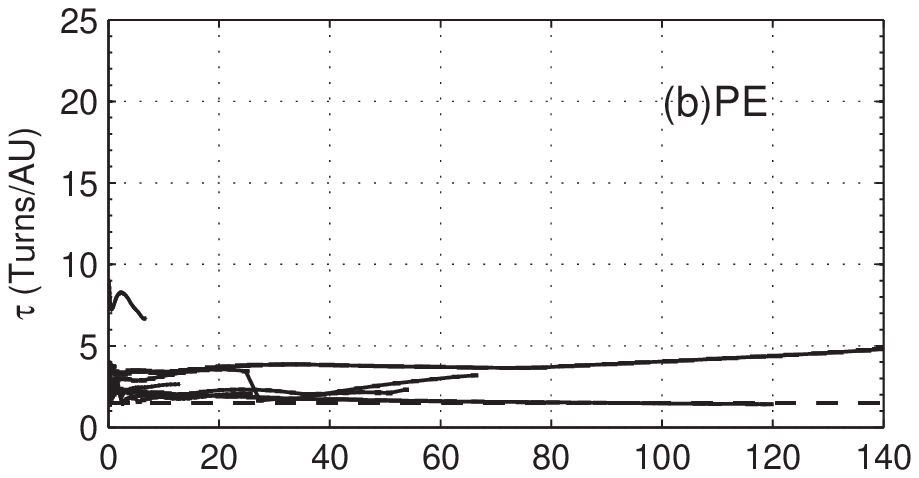}\\
\includegraphics[]{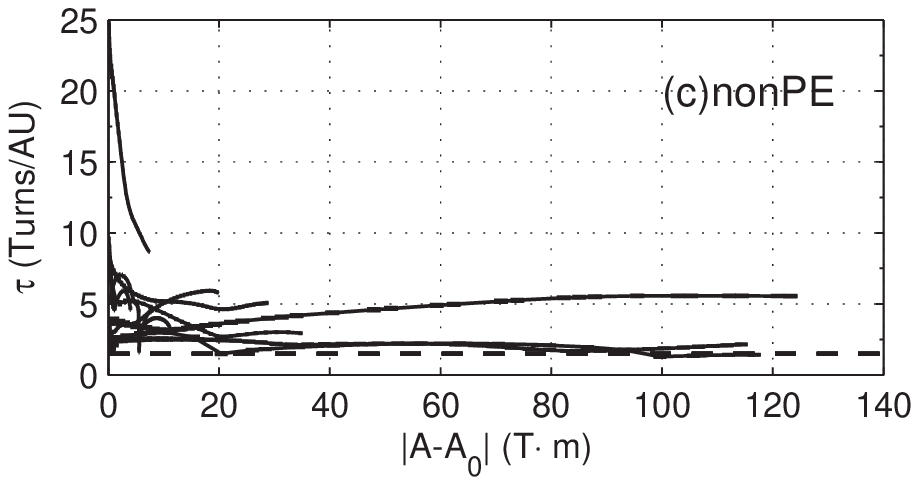}\\
\caption{Field-line twist ($\tau$, turns/AU) distribution along
$A$ shells (i.e., vs. $A'$) for (a) all events, (b) the ones
associated with prominence eruption, and (c) the ones without
prominence eruption. The horizontal dashed line is of value 1.5.}
\label{figtauA}
\end{figure}

\begin{figure}
\includegraphics[]{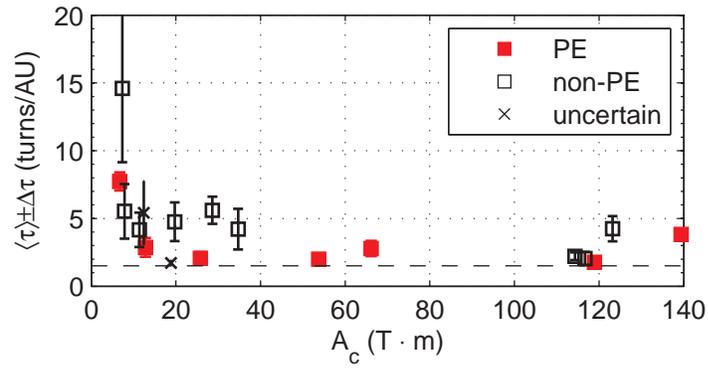}
\caption{The average and standard deviation (error bars) of the
field-line twist $\tau(A)$ for all events. {The horizontal axis
denotes the cut-off boundary $A'\equiv A_c$ (the maximum value of
$A'$ of each line in Figure~\ref{figtauA}) within which the
graphic method of determining the field-line twist works.} The
events associated with prominence eruption are marked with filled
symbols.} \label{figtauAc}
\end{figure}

\begin{figure}
\includegraphics[width=1.0\textwidth]{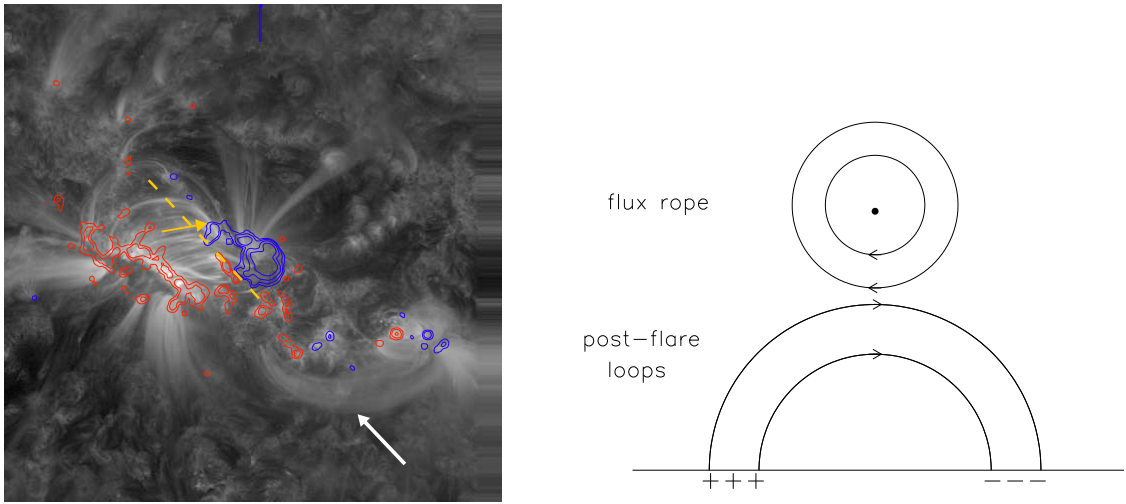}
\caption{Left: snapshot of the post-flare loops observed in
171\AA\ by AIA, with superimposed contours of the longitudinal
magnetogram by HMI. Red (blue) contours denote positive (negative)
magnetic fields and the contour levels indicate the field strength
at $\pm$ 100, 200, 400, and 800~G. The orange dashed line marks
the magnetic PIL in this event, and the orange arrow shows the
direction of magnetic field at the flare loop top based on the
morphology of the post-flare loops. Right: sketch of the
cross-section of reconnection formed flux rope and post-flare
loops below it, as viewed from the { southwest along the PIL
demonstrated by the white arrow in the left panel.} Solid lines
with arrows indicate magnetic fields of these structures, and the
"$\cdot$" sign in the middle of the flux rope indicates the
outward direction of the axial magnetic field. }
\label{fighelsign}
\end{figure}

\begin{figure}
% \vspace*{-2.0 cm}
\centering
 \includegraphics[width=.45\textwidth]{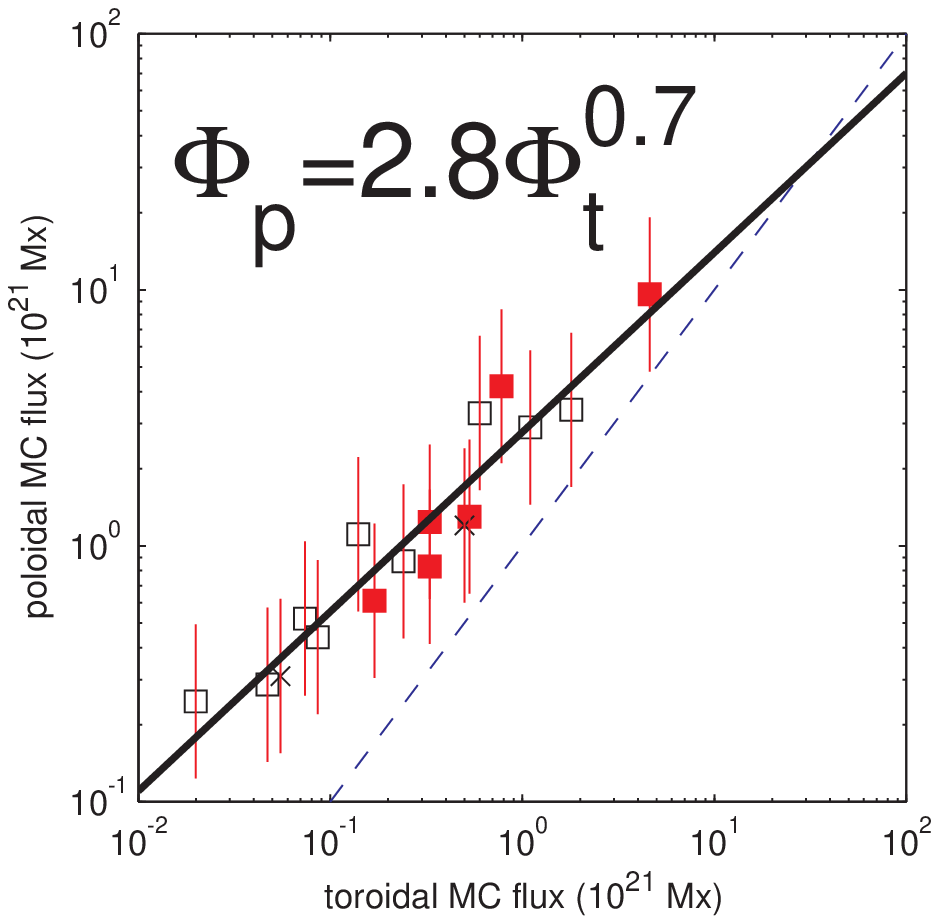}
 \includegraphics[width=.45\textwidth]{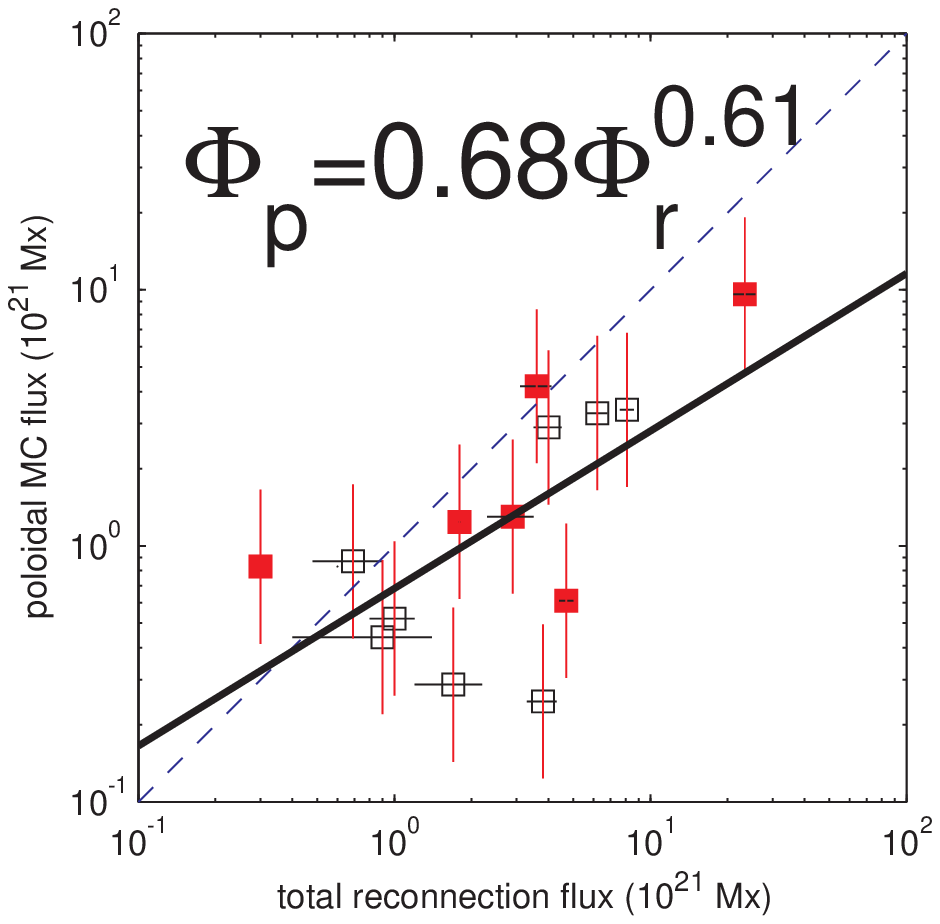}
% \vspace*{-1.0 cm}
 \caption{ Magnetic flux comparison of poloidal flux $\Phi_p$ vs. toroidal flux
$\Phi_t$ ({\em left}) and poloidal flux $\Phi_p$ vs. reconnection
flux $\Phi_r$ ({\em right}). The events associated with prominence
eruption (P.E.) are marked by filled squares. {The ones with
uncertain P.E. association (events \#12 and \#14) are marked by
the cross symbols and excluded in the right panel. The
least-squares fit
 to each data set is given and illustrated by the thick solid line. The dashed line indicates the one-to-one line. The correlation coefficients are 0.95 and 0.63,
 respectively.}}
   \label{figflux}
\end{figure}

\begin{figure}
\includegraphics[width=.6\textwidth]{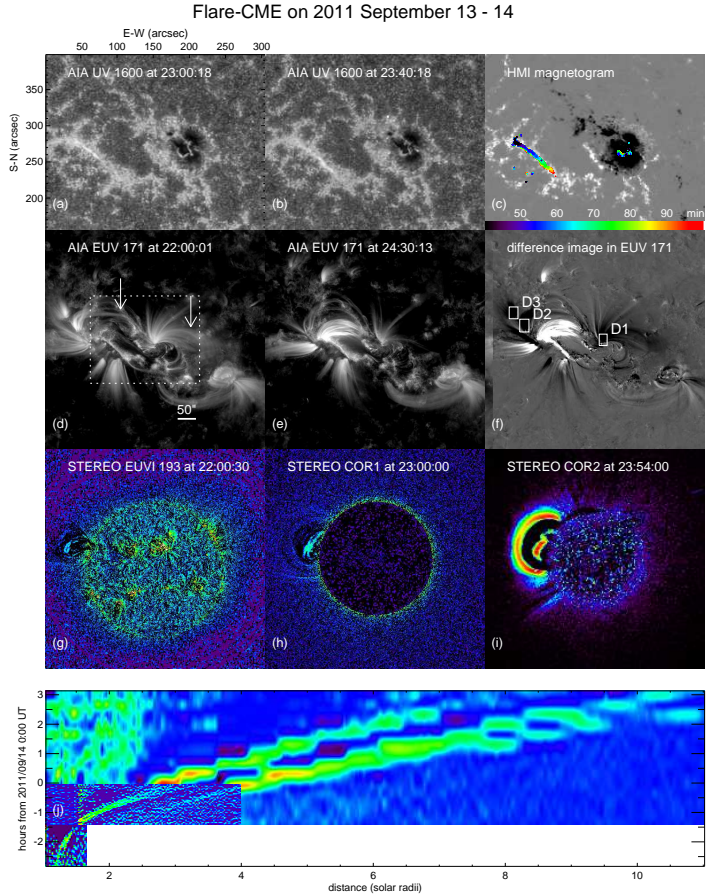}
\caption{\small Evolution of flare ribbons observed in UV 1600\AA\
images by AIA (a, b), pre- and post-flare loops observed in EUV
171\AA\ images (d, e), and evolution of the associated CME
observed by EUVI (g), COR1 (h), COR2 (i) onboard STEREO-A on 2011
September 13-14. The time sequence of ribbon brightening is mapped
onto the longitudinal magnetogram by HMI in panel (c). { The field
of view (FOV) of the images in panels (d) - (f) is larger than
that in panels (a) - (c), and the dotted box in panel (d)
indicates the FOV of the images in (a) - (c). The heliographic
coordinates of these images are given in panel (a).} Arrows in
panel (d) indicate pre-flare sheared loops that are disrupted
during the flare. Panel (f) is a difference image of (d) and (e),
and the three boxes denote the regions where dimming is observed
and analyzed, which is shown in Figure~\ref{fig0914_2}. The bottom
panel shows the time-distance intensity-gram constructed using
data from EUVI, COR1, and COR2 along a straight slit connecting
solar center and the top of the erupting CME.} \label{fig0914_1}
\end{figure}

\begin{figure}
\centering
\includegraphics[width=1.\textwidth]{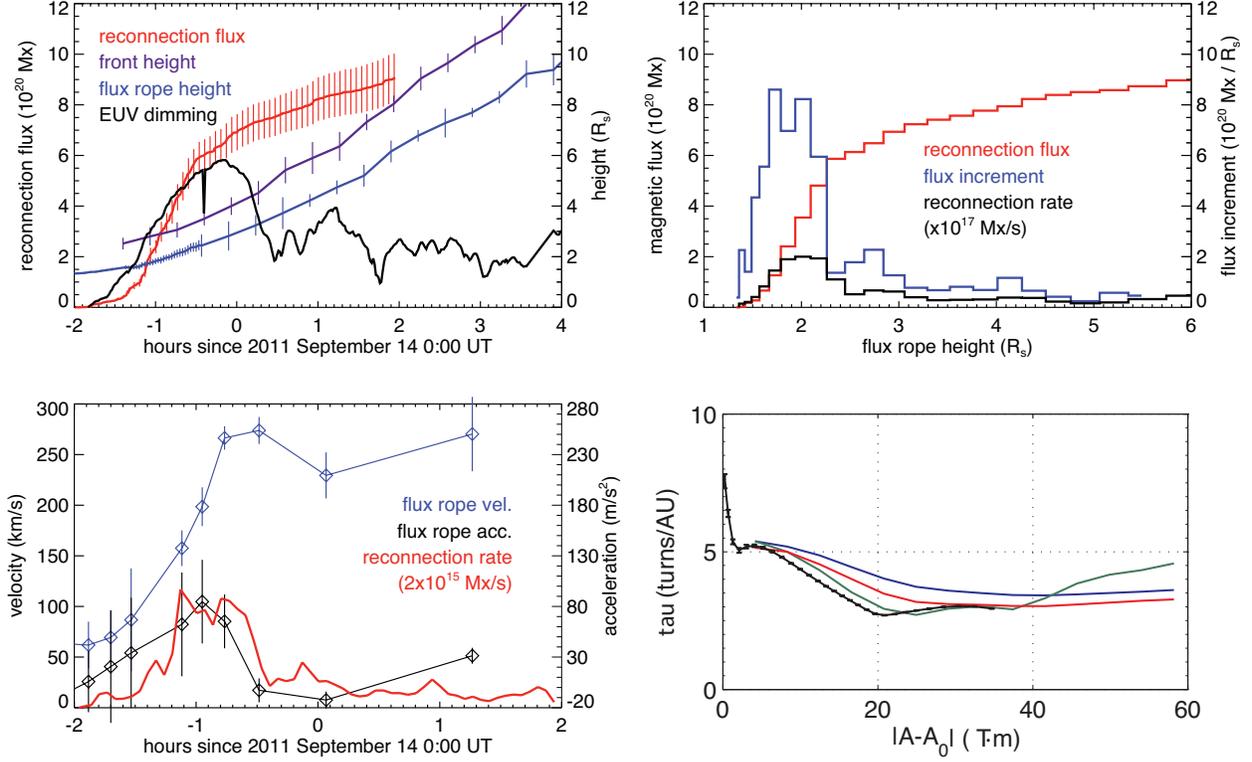}
\caption{Top left: time sequences of the heights of the CME core (blue)
and front (violet) measured in STEREO images in comparison with
flare reconnection flux (red) measured in UV images and inverted
EUV 171\AA\ intensity of the active region (black) showing
the occurrence of dimming followed by formation of bright post-flare
loops. Lower left: velocity (blue) and acceleration (black) of the
CME core in comparison with reconnection rate (red). Top right:
reconnection flux (red), flux increment (blue; see text), and
reconnection rate (black) versus the height of the CME core. Lower
right: twist of the MC field lines as a function of $A'$,
indicating twist distribution from the core outward. Different
colors show measurements with four different methods (black line
is from the graphic method; see the Appendix). } \label{fig0914_2}
\end{figure}

\begin{figure}
\includegraphics[width=.8\textwidth]{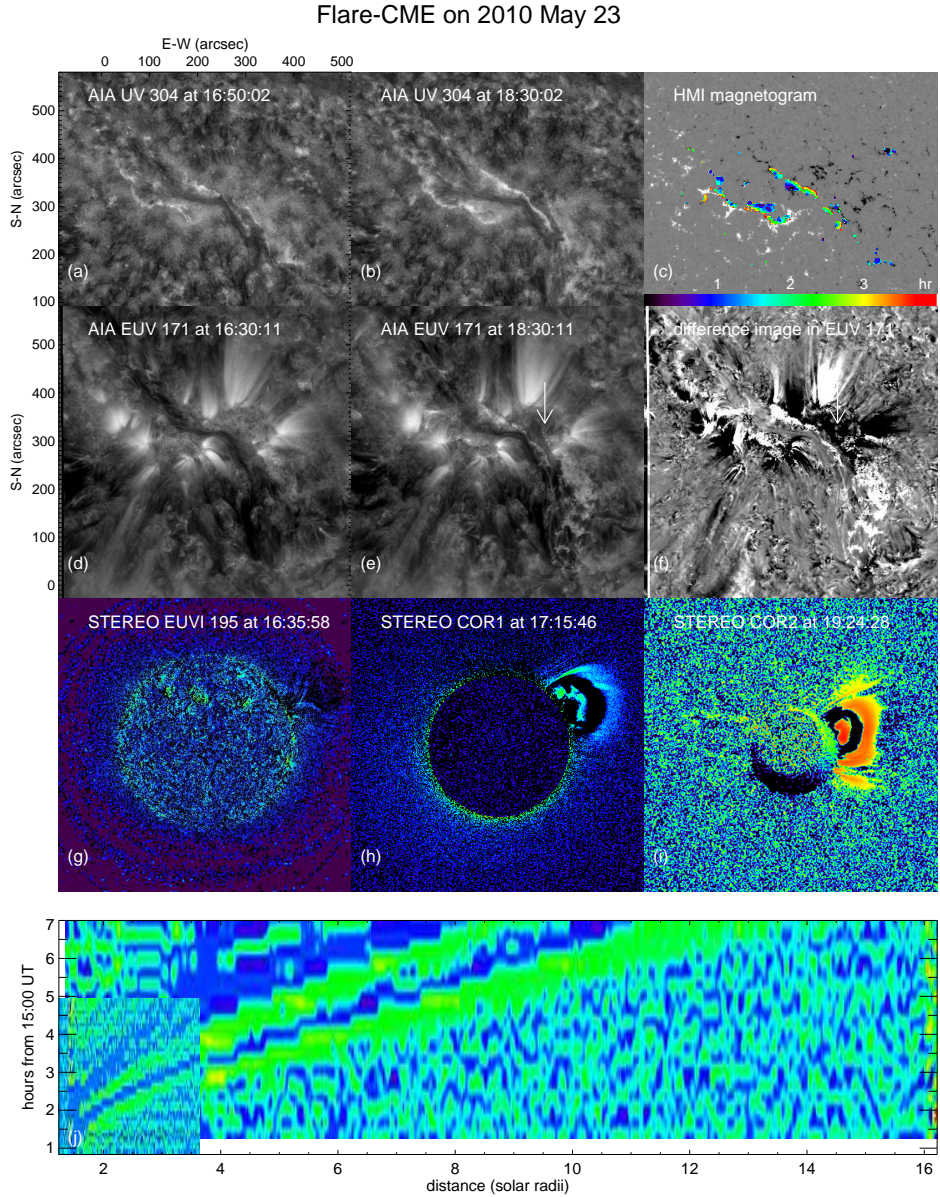}
\caption{Same as Figure~\ref{fig0914_1} but for the flare/CME
event observed on 2010 May 23. The top panels show flare ribbons
observed in EUV 304\AA\ by AIA. Arrows in the middle panels
indicate the dimming patch next to the ribbons. In the bottom
panel, the time-distance intensity-gram is constructed using only
COR1 and COR2 data because the CME is not well observed in EUVI
images. } \label{fig0528_1}
\end{figure}

\begin{figure}
\includegraphics[width=1.\textwidth]{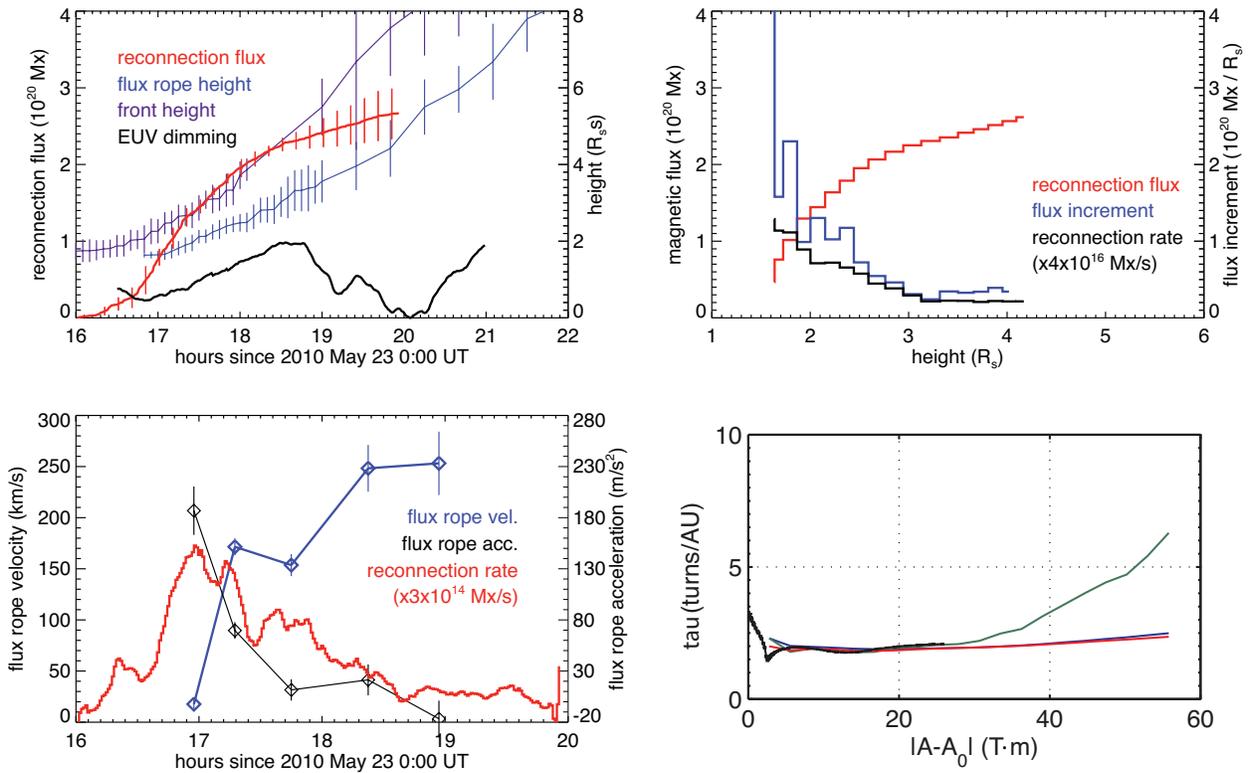}
%\includegraphics[width=.45\textwidth]{accl_0914.eps}
%\hspace{.8cm}
%\includegraphics[width=.45\textwidth]{fig_twist2.eps}
\caption{Same as Figure~\ref{fig0914_2} but for the flare/CME/MC
event on 2010 May 23-28.} \label{fig0528_2}
\end{figure}

\begin{figure}
\begin{minipage}[b]{.6\textwidth}
\includegraphics[width=1.\textwidth]{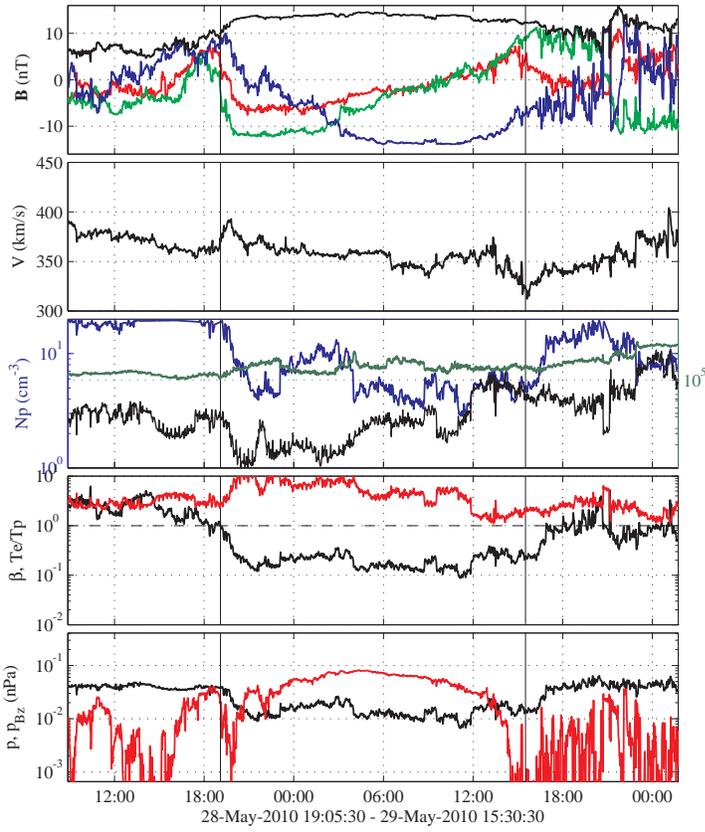} \centering(a)
\end{minipage}
%\caption{} \label{fig0914_data}
%\end{figure}
%\begin{figure}
\begin{minipage}[b]{.4\textwidth}\centering
(b)\includegraphics[width=1.\textwidth]{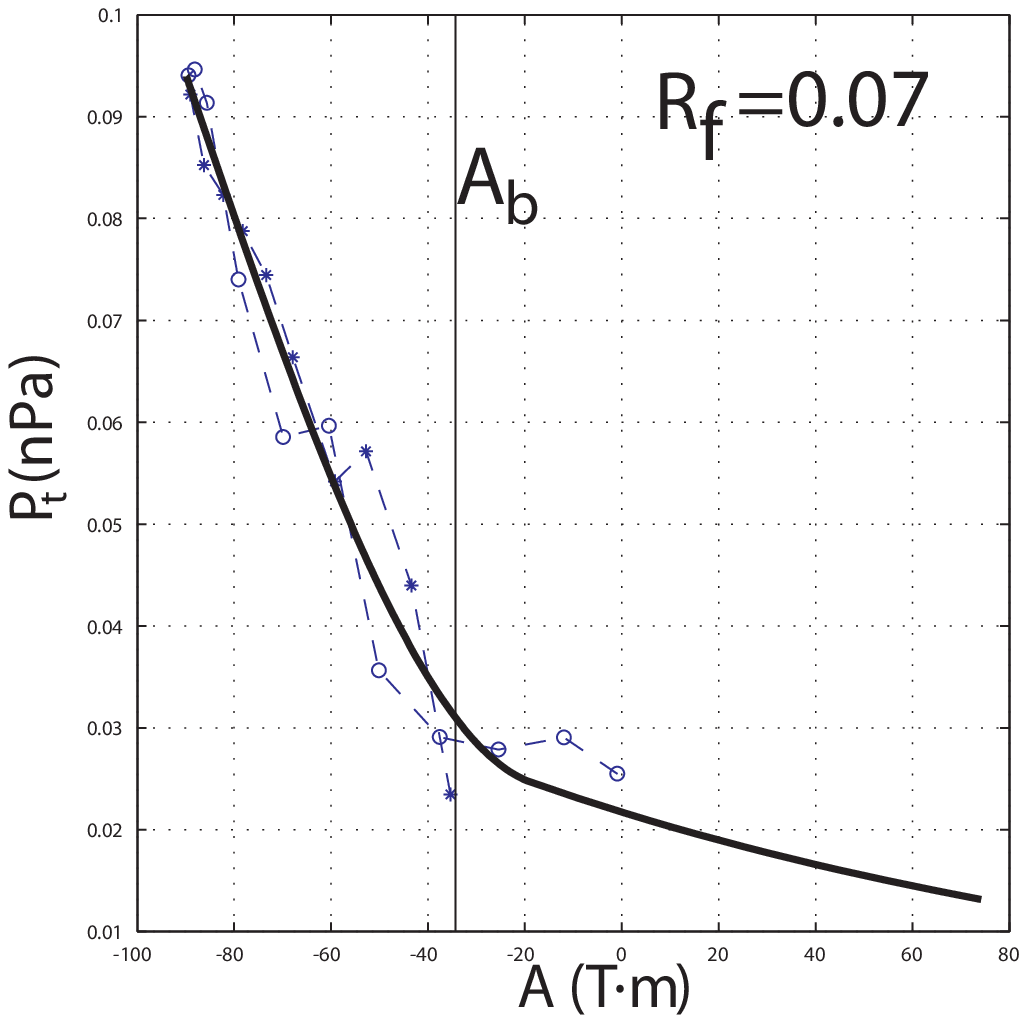}
(c)\includegraphics[width=1.\textwidth]{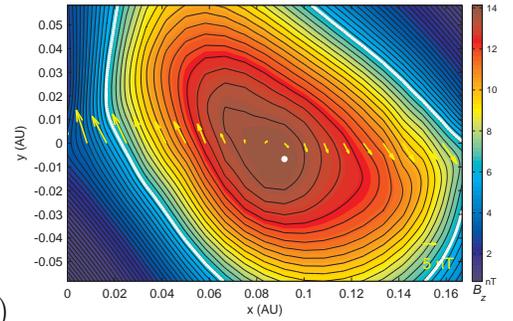}
\end{minipage}
\caption{The basic GS reconstruction result for event \#11 from
Wind spacecraft in-situ measurements. The format is the same as
Figure~\ref{fig0914_ALL}.} \label{fig0528_all}
\end{figure}

\begin{figure}
\includegraphics[]{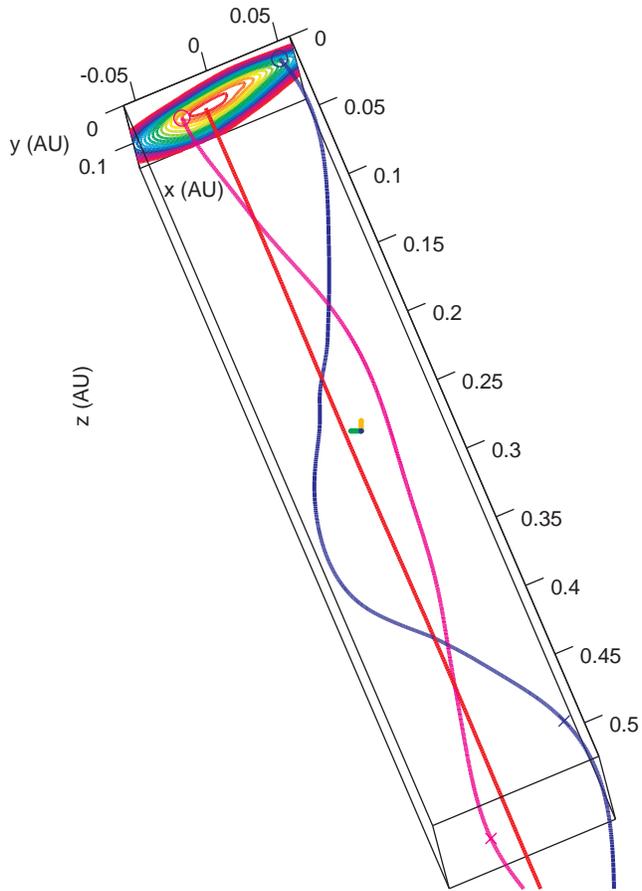}
\caption{A 3D view toward the Sun of the flux-rope configuration of
event \#11. The format is the same as Figure~\ref{fig0914_3D}.}
\label{fig0528_3D}
\end{figure}

The graphic method, by finding the axial length, $L_z$ in AU, of
each field line completing one turn along each $A$ shell, yields
the exact field-line twist
\begin{equation}
\tau(A)=\frac{1}{L_z(A)}, \label{eq:tau}
\end{equation}
in unit of turns/AU. Because this method requires a field line to
complete one turn in the computational domain (an elongated box in
Figures~\ref{fig0914_3D} and \ref{fig0528_3D}), it only applies to
closed contours of $A$ in the cross-sectional map of GS
reconstruction result. In other words, the range of valid $A$
values is limited and there is a cut-off $A$ value, denoted $A_c$,
corresponding to the outermost loop of $A$ from the center of the
flux rope, beyond which this method does not apply. This cut-off
boundary also has implications for the other methods. As we
demonstrate below, the accuracy of the approximation  degrades
greatly beyond this boundary.

To test the validity of the various field-line twist estimates
described above, we apply those methods to two analytic flux-rope
models with known field-line twist distributions. One is the
so-called Gold-Hoyle (GH) model with constant field-line twist
\citep{1960MNRAS.GH,1999AIPCF}. The other is the well-known
linear-force free Lundquist flux-rope model \citep{lund}. Both have
axi-symmetric cylindrical geometry, i.e., everything is
dependent on the radial distance $r$ from the flux-rope center
only. We construct the two models based on real events which
provide the necessary fitting parameters: the maximum axial
magnetic field $B_0$, and the size (maximum radius $R_0$) of the
flux rope within which the axial field $B_z$ remains unipolar. The
axial and azimuthal magnetic field components of the GH model are
given as \citep[e.g.,][]{1999AIPCF}
\begin{eqnarray}B_z&=&\frac{B_0}{1+T_0^2 r^2}\label{eqGH1}\\
B_\phi&=&\frac{B_0 T_0 r}{1+T_0^2 r^2},\label{eqGH2}\end{eqnarray}
where a constant field-line twist is written $T_0$ of unit
radians/AU which differs by a factor $2\pi$ from our definition of
field-line twist in the unit of turns/AU.

\begin{figure}
\includegraphics[width=0.46\textwidth]{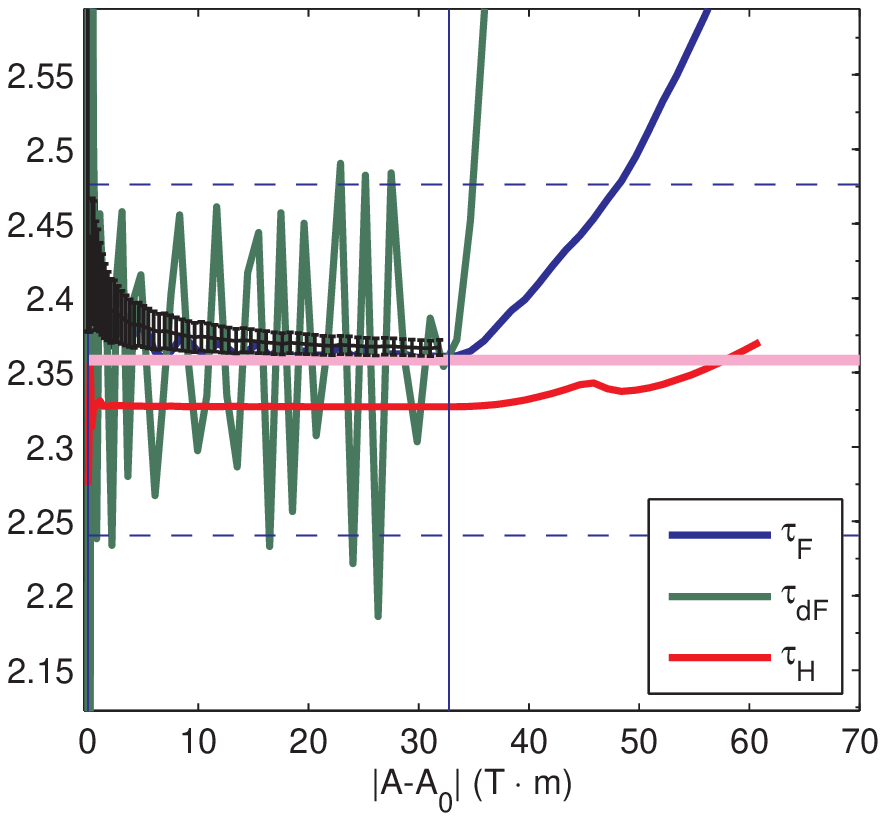}
\includegraphics[width=0.45\textwidth]{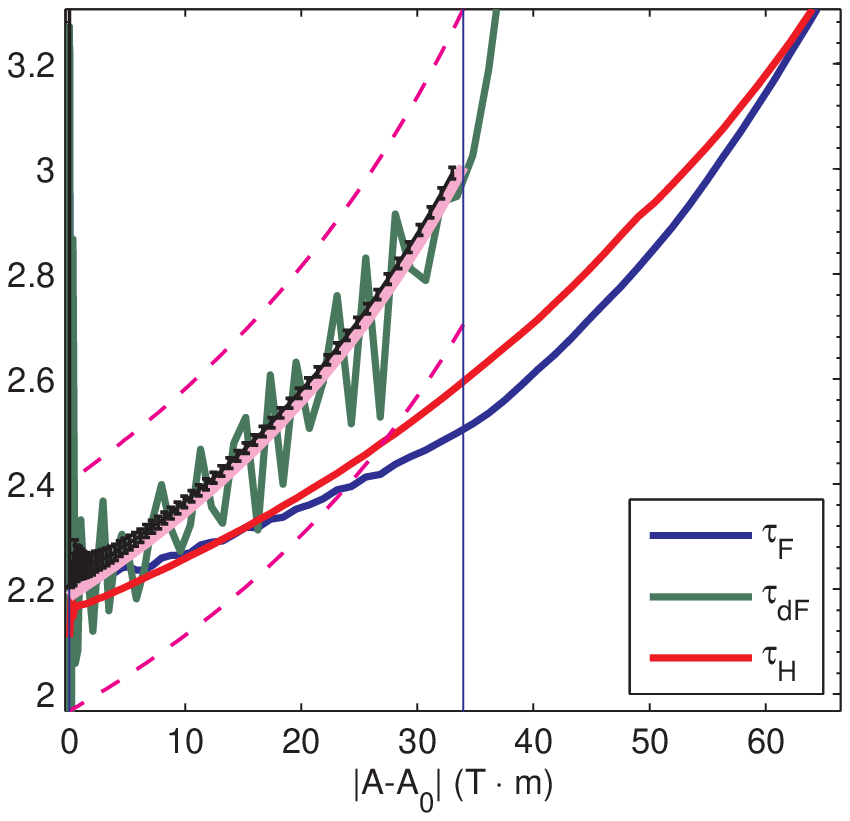}
\caption{Test case of field-line twist  estimates (all in
turns/AU) for a relatively large-size flux rope: (left panel)
constant twist Gold-Hoyle flux-rope model, and (right panel) the
linear force-free Lundquist flux-rope model. Black line with error
bars is the estimate obtained by the graphic method. Thick pink
line is the exact value. The thin vertical line denotes the
cut-off boundary $A_c$. In the left (right) panel, the dashed line
indicates a 5\% (10\%) uncertainty zone. } \label{figFLtest}
\end{figure}

Two cases were examined. For each case, both the GH model and the
Lundquist model were constructed. Figure~\ref{figFLtest} shows the
case of a relatively large-size flux rope with large  $A$ values
and $B_0=14$ nT, $R_0=0.088$ AU for the Lundquist model.
Figure~\ref{figFLtest} (left panel) shows the GH model with a
constant twist $T_0=15$ radians/AU as indicated by the thick pink
line. All the approximations fall within the 5\% uncertainty zone
around the exact value for the closed contour region where
$|A-A_0|\le A_c$. The green line shows a great deal of fluctuation
due to the nature of the finite-difference approximation in
Equation~(\ref{eq:tauF2}). All results diverge at the center of
the flux rope where the field line simply becomes straight so that
the twist value is arbitrary. Beyond the closed loop boundary as
marked by the vertical line, the approximations begin to
deteriorate greatly because there are no more closed $A$ contours
present in the rectangular solution domain. Therefore the twist
estimate from the graphic method ceases to exist beyond this
boundary. Both the toroidal flux and helicity contents also become
under-estimated which leads to large deviations from the exact
twist value beyond the cut-off boundary. The estimate $\tau_{dF}$
(green curve) starts to increase rapidly due to the significant
reduction of $\Delta\Phi_t$ while $\Delta\Phi_p$ remains properly
evaluated.

\begin{figure}
\includegraphics[width=0.47\textwidth]{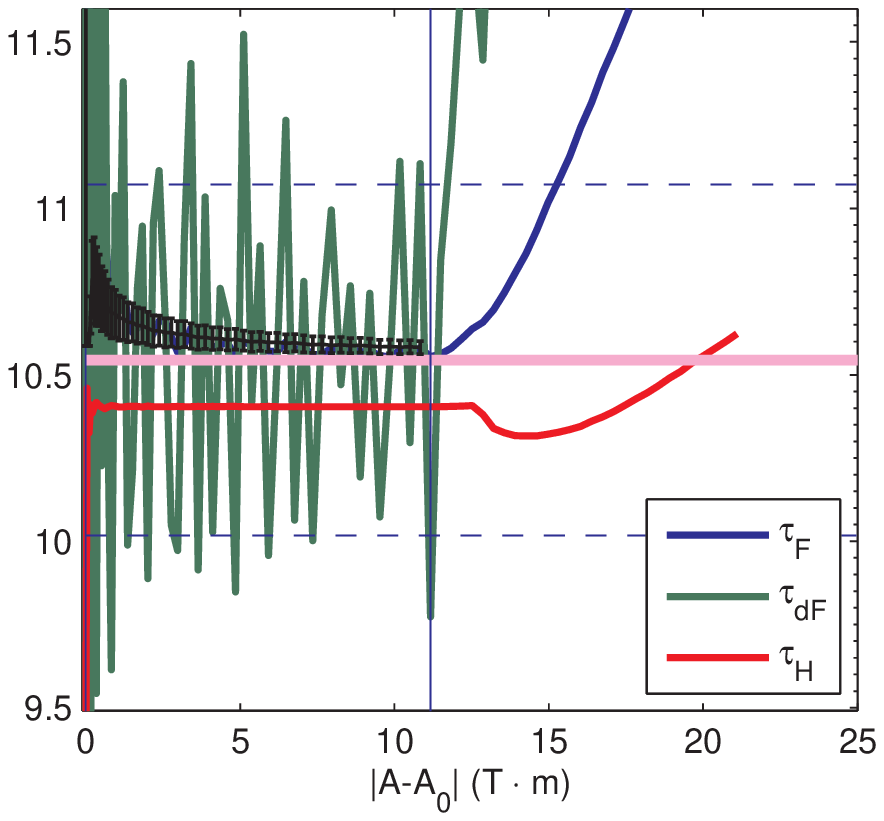}
\includegraphics[width=0.45\textwidth]{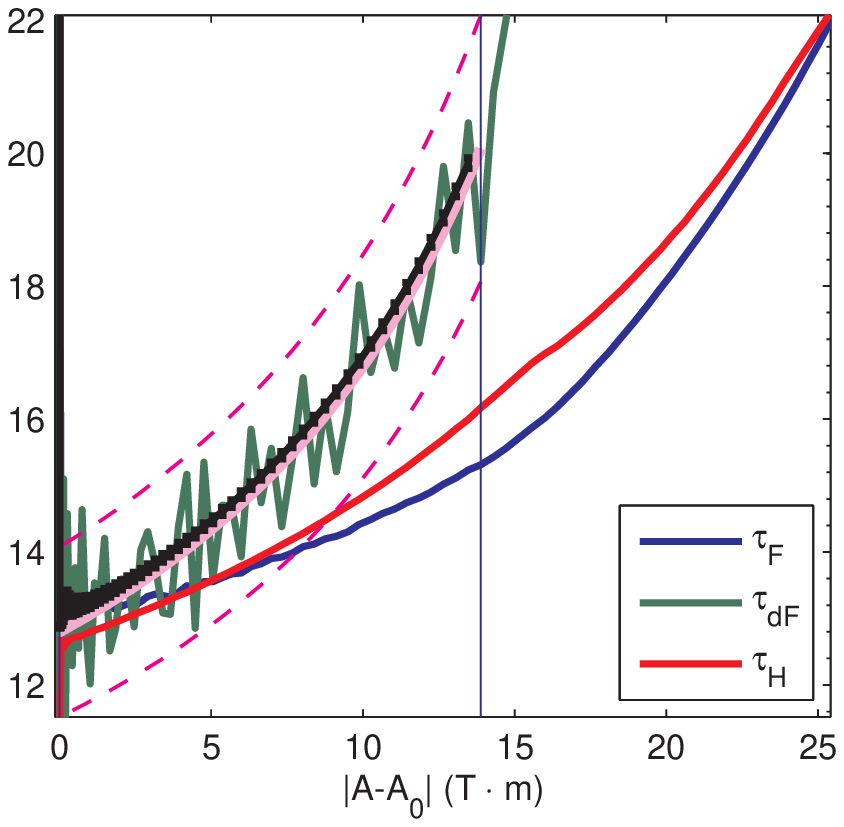}
\caption{Test case of field-line twist  estimates for a relatively
small-size flux rope. Format is the same as
Figure~\ref{figFLtest}.} \label{figFLtest2}
\end{figure}

Figure~\ref{figFLtest} (right panel) shows the corresponding
results for the Lundquist model where the exact field-line twist
increases monotonically from the center of the flux rope toward
the boundary. The result from the graphic method again follows the
exact result most closely, while the green line fluctuates around
within a 10\% bound and the other two show greater deviations.
Compared with Figure~\ref{figFLtest} (left panel), the three
flux/helicity based approximation methods perform worse in this
case of a non-constant twist. For completeness, we also examined a
second case of a relatively small size but significantly larger
twist, with $B_0=27$ nT, $R_0=0.015$ AU for the Lundquist model,
and $T_0=66$ radians/AU for the GH model. Figure~\ref{figFLtest2}
 shows the field-line twist estimates in the same format as
Figure~\ref{figFLtest} for the GH flux-rope model and the
Lundquist flux-rope model, respectively. The behavior of these
results is very similar to what we have discussed for
Figure~\ref{figFLtest}, despite the significant difference in size
and twist value.

Overall we conclude that the graphic method is clearly the most
reliable method for estimating the magnetic field-line twist along
the $A$ shell. The results are available and valid for the region
surrounding the center of the flux rope where $|A-A_0|\le A_c$
excluding the center where $A\equiv A_0$. Our test results also
indicate that all other approximations are more likely to provide
results consistent with the graphic method for a flux-rope
configuration of constant field-line twist. Among them the
estimate $\tau_{dF}$ agrees well with  $\tau$, especially for a
varying twist distribution within the cut-off boundary. All these
estimates (Equations~\ref{eq:tauH}-\ref{eq:tauF2}) are based on
magnetic flux and helicity contents which are well conserved
quantities in ideal MHD.  We base our discussions in the main body
of the text mostly on these conclusions.

%\bibliographystyle{apj}
%\bibliography{ref_master}
\bibliographystyle{apj}
\bibliography{ref_master}

%\input{table_rev2}
%\clearpage
% the following is the information table for event #10-19; see Qiu et al. (2007) for event 1-9
\begin{deluxetable}{ccccccccccc}
\tabletypesize{\scriptsize}
\rotate \tablecolumns{11} \tablewidth{0pt} \tablecaption{Event
Information\tablenotemark{a}\label{info_list}}
    \tablehead{
    \colhead{Event\#}&
    \multicolumn{3}{c}{Flare\tablenotemark{b}} &
    \colhead{Filament} &
    \multicolumn{2}{c}{CME\tablenotemark{d}} &
    \colhead{} &
    \multicolumn{3}{c}{Magnetic Cloud/ICME\tablenotemark{e}} \\
    \cline{2-4} \cline{6-7} \cline{9-11}\\
    \colhead{} &
    \colhead{Date} &
    \colhead{Region} &
    \colhead{Time \& Mag.} &
    \colhead{Info. \tablenotemark{c}} &
    \colhead{Time} &
    \colhead{Speed} &
    \colhead{} &
    \colhead{Date} &
    \colhead{GS Interval (mm/dd/yyyy hh:mm:ss)} &
    \colhead{Duration (hrs)}}

\startdata

10 & 2008 & - & -     & yes & - & - & & Mar 08& 03/08/2008 18:42:41-03/09/2008 01:07:45 & 6.4 \\
11 & 2010 May 23 & N19W12 & 16:30 B1.1 & yes & 18:06 & 258 & & May 28 & 05/28/2010 19:05:30-05/29/2010 15:30:30 & 20 \\
12 & 2010 Aug 01 & N20E36 & 07:24 C3.2 & uncertain & uncertain & uncertain & & Aug 04& 08/04/2010 03:53:00-08/04/2010 07:47:40 & 3.9 \\
13 & 2011 Feb 15 & S20W12 & 02:00 X2.2 & no & 02:24 & 669 & & \nodata & \\% & -  & -   & - \\
14 & 2011 Mar 25 & S16E30 & uncertain & uncertain & uncertain & uncertain & & Mar 30 & 03/30/2011 03:17:11-03/31/2011 14:55:51 & 36 \\
15 & 2011 Jun 02 & S18E28 & 07:22 C2.2 & no & 08:12 & 976 & & Jun 05 & 06/05/2011 01:17:32-06/05/2011 06:29:00 & 5.2 \\
16 & 2011 Aug 04 & N16W38 & 03:41 M9.1 & no & 04:12 & 1315 & & Aug 05 & 08/05/2011 20:05:55-08/05/2011 22:03:15 & 2.0 \\
17 & 2011 Sep 04 & N19W87 & 23:58 C7.9 & no & (+1)00:48 & 622 & & Sep 09 & 09/09/2011 01:20:00-09/09/2011 12:28:00 & 11 \\
18 & 2011 Sep 13 & N23W21 & 22:30 C2.9 & no  & (+1)00:05 & 408 & & Sep 17 & 09/17/2011 15:18:52-09/18/2011 05:08:44 & 14 \\
19 & 2011 Oct 22 & N30W30 & - & yes  & 01:25 & 593 & & Oct 24 & 10/24/2011 22:16:53-10/25/2011 13:18:13 & 15 \\
\enddata
\tablenotetext{a}{References for event identification and
association of the MC, CME,  flare and filament eruption are
listed below. References are not given
for event \#13 since the MC was not successfully analyzed by the GS method. See Section 2 for detailed explanation of the identification by various sources. \\
10: \citet{Li2014} \\
11: \citet{Li2014, Lugaz2012, Mostl2014}\\
12: \citet{Li2014, Torok2011, Schrijver2011, Titov2012, Harrison2012, Mostl2012, Mostl2014}; See Section~2 for details. \\
14: \citet{Li2014, Savani2013}\\
15: \citet{Li2014} \\
16: \citet{Gopal2012} \\
17: In this event, the Magnetic Cloud (GS Interval) is measured at
the STEREO-A spacecraft, and its association with CME and flare is
verified by private communication with Dr. C. C. Wu and further
examined by authors with the STEREO EUVI movies. Since
the MC/CME is associated with a limb flare, magnetic reconnection flux is not measured in this event.\\
18: \citet{Li2014} \\
19: \citet{Li2014, Mostl2014}\\
In addition, association between MC and CME of events \#11, 12,
16, 18 is also provided in the online catalogue ``Near-Earth
Interplanetary Coronal Mass Ejections Since January 1996" compiled
by I. Richardson and H. Cane at
$http://www.srl.caltech.edu/ACE/ASC/DATA/level3/icmetable2.htm$,
and association between MC, CME, and solar surface activities
including flares of events \#11, 12, 14, 16, 18, 19 is also
provided in the online catalogue ``GMU CME/ICME List" compiled by
Phillip Hess and Jie Zhang at
$http://solar.gmu.edu/heliophysics/index.php/GMU\_CME/ICME\_List$.
}

\tablenotetext{b}{Information is obtained from
$http://solarmonitor.org/$. Time refers to the start time of GOES
X-ray flux increase, and magnitude refers to GOES categorization.}

\tablenotetext{c}{``yes" indicates filament eruption detected, and ``no" indicates filament eruption not seen.}

\tablenotetext{d}{Information is obtained from $http://cdaw.gsfc.nasa.gov/CME\_list/$. Time refers to when the
CME is first observed in LASCO C2 field of view (FOV), and speed, in units of km s$^{-1}$, refers to the linear fit to the
height-time profile obtained from C2-C3 observations.}

\tablenotetext{e}{The intervals were identified and utilized based
on GS reconstruction of magnetic flux ropes embedded within each
ICME complex, which do not necessarily coincide with the intervals
identified by other criteria. All intervals correspond to in-situ
measurements at Earth except for event \#17 which is at STEREO-A.}
\end{deluxetable}

\setlength{\tabcolsep}{0.02in}

\begin{deluxetable}{ccccccccccc}
\tabletypesize{\scriptsize} \rotate \tablecaption{Master table of
relevant results for all events\label{table2}} \tablewidth{0pt}
\tablehead{ \colhead{Event\#} &
\colhead{$(\Phi_r\pm\Delta\Phi_r)$} & \colhead{Pattern} &
\colhead{Sign of Helicity} & \colhead{$B_{z0}$} &
\colhead{$\Phi_{t,max}$} & \colhead{$\Phi_{p,max}$} &
\colhead{$K_{r,max}$} & \colhead{$\bar\tau_H$}&
\colhead{$\bar\tau_F$}&
\colhead{Mean twist} \\
% now the second row of the header
\colhead{} & \colhead{$10^{21}$ Mx} & \colhead{} &
\colhead{Flare/MC\tablenotemark{a}} & \colhead{nT} &
\colhead{$10^{21}$ Mx} & \colhead{$10^{21}$ Mx} &
\colhead{$10^{42}$ Mx$^2$} &
\colhead{$\frac{K_{r,max}}{\Phi_{t,max}^2}$} &
\colhead{$\frac{\Phi_{p,max}}{\Phi_{t,max}}$} &
\colhead{$(\langle\tau\rangle\pm\Delta\tau)$ } }

\startdata
1 & (4.0$\pm$0.5) & U & R/R  &    26   & 1.1 & 2.9  & 3.2& 2.4& 2.5& (2.2$\pm$0.28) \\
2 & (1.0$\pm$0.2)& $\perp$ & L/L  & 14    &    0.074 & 0.52 & 0.028 & 5.2& 7.0 & (4.2$\pm$1.3) \\
3 &(2.9$\pm$0.6) & $\perp$ & L/L &    32 &      0.53& 1.3 & 0.75 & 2.6& 2.5& (2.8$\pm$0.56) \\
4 & (4.7$\pm$0.3)& $\perp$ & R/R&     21  &      0.17& 0.61 & 0.092 & 3.1& 3.5&(2.9$\pm$0.69) \\
5 &(0.9$\pm$0.5) & $\perp$ & L/L   &  11 &     0.086 & 0.44 & 0.037 & 5.1& 5.2&(4.8$\pm$1.4) \\
6 &(23.4$\pm$2.3) & $||$ $\perp$ & L/L  &     45    &     4.6 & 9.6 & 35    & 1.6 & 2.1 &(1.8$\pm$0.28) \\
7 &(3.6$\pm$0.5) & $\perp$ & U/R  &    56    &   0.78& 4.2    & 3.2    &   5.2& 5.4&(3.8$\pm$0.37) \\
8 &(6.2$\pm$0.6) & $||$ $\perp$ & L/L  &  39  &    0.60& 3.3& 2.1& 5.8& 5.6&(4.2$\pm$0.93) \\
9 &(8.1$\pm$0.5) & $||$ $\perp$ & L/L  & 54  &    1.8   &    3.4   &    4.9 &      1.5  &     1.9 &  (2.0$\pm$0.49) \\
10 &- & - & -/R &   10 &     0.018& 0.14 & 0.0024 & 7.3& 7.6&(7.7$\pm$0.67) \\
11 &(0.3$\pm$0.3) & $\perp$ & L/L &    14   &    0.33& 0.83    & 0.26 & 2.4& 2.5&(2.0$\pm$0.46) \\
12\tablenotemark{b} & (0.83$\pm$0.38)  & $\perp$ & L/R &    17  &      0.055& 0.31  & 0.016  & 5.2& 5.6&(5.4$\pm$2.4) \\
{\em 13} &(6.7$\pm$0.6) & \nodata& \\
14\tablenotemark{b} & (0.42$\pm$0.03) & U & R/R &  13  &   0.50& 1.2& 0.51   & 2.0& 2.5&(1.7$\pm$0.22) \\
15 &(1.7$\pm$0.5) & $||$ $\perp$ & R/R &    20   &    0.047& 0.29 & 0.014& 6.5& 6.1&(5.5$\pm$2.0) \\
16 &(3.8$\pm$0.5) & U & U/R &     27 &      0.020& 0.25 & 0.0042 & 11& 12&(14.6$\pm$5.4) \\
17 &- & - & - &    18   &      0.14 & 1.1 & 0.16 & 8.1 & 7.9 &(5.6$\pm$1.0) \\
18 &(0.69$\pm$0.21) & $||$ & L/L & 13  &    0.24 &0.87& 0.19    & 3.3 &3.6&(4.2$\pm$1.5) \\
19 &- & - & L\tablenotemark{c}/L &     24 &       0.44& 0.93 & 0.42 & 2.2& 2.1&(2.0$\pm$0.27) \\
\enddata
\tablenotetext{a}{L: left-handed; R: Right-handed; U:
Undetermined.}

\tablenotetext{b}{For event \#12, the reconnection flux is
measured for the C3.2 flare. For event \#14, the reconnection flux
is measured in the M1.0 flare, only as an upper-limit of
reconnected flux possibly associated with the MC, and the sign of
helicity is determined from morphology of flares in the same
active region. See Section 2 for details. Both are excluded from
the flux comparison of $\Phi_p$ vs. $\Phi_r$.}

\tablenotetext{c}{Sign of helicity determined from the filament.}
\end{deluxetable}

\clearpage

%% This figure uses \includegraphics to scale and rotate the still frame


\end{document}